\documentclass[aps,prb,twocolumn,showpacs,superscriptaddress]{revtex4}
\usepackage{graphicx}
\usepackage{amsmath}
\usepackage{amstext}
\usepackage{amssymb}
\usepackage{mathrsfs}
\usepackage{amsfonts}
\usepackage{amsbsy} 
\usepackage{dcolumn}
\usepackage{bm}
\usepackage{multirow}
\usepackage{soul}
\usepackage{datetime}

\newcommand{\bea}{\begin{eqnarray}}
\newcommand{\eea}{\end{eqnarray}}
\newcommand{\E}{\mathop{\mathrm{E}}}
\newcommand{\q}{\mathbf{q}}
\newcommand{\U}{\text{U}}
\newcommand{\GSD}{\mathop{\mathrm{GSD}}}

\makeatletter
\newcommand{\xRightarrow}[2][]{\ext@arrow 0359\Rightarrowfill@{#1}{#2}}
\makeatother

\begin{document}
\title{Symmetry-protected topological phases with charge and spin symmetries:\\
response theory and dynamical gauge theory in 2D, 3D and the surface of 3D}
\author{Peng Ye}  
\affiliation{Perimeter Institute for Theoretical Physics, Waterloo, ON, N2L 2Y5, Canada}
\author{Juven Wang}
\affiliation{Department of Physics, Massachusetts Institute of Technology, Cambridge, MA 02139, USA}
\affiliation{Perimeter Institute for Theoretical Physics, Waterloo, ON, N2L 2Y5, Canada}
\begin{abstract}
A large class of symmetry-protected topological phases (SPT) in boson / spin systems have been recently predicted by the group cohomology theory.  In this work, we consider SPT states {\it at least} with charge symmetry (U(1) or Z$_N$) or spin $S^z$ rotation symmetry (U(1) or Z$_N$)  in 2D, 3D, and the surface of 3D. If both are U(1), we apply external electromagnetic field / ``spin gauge field'' to study the charge/spin response.  For the SPT examples we consider (\emph{i.e.} U$_c$(1)$\rtimes$Z$^T_2$, U$_s$(1)$\times$Z$^T_2$, U$_c$(1)$\times$[U$_s$(1)$\rtimes$Z$_2$]; subscripts $c$ and $s$ are short for charge and spin; Z$^T_2$ and Z$_2$ are time-reversal symmetry and $\pi$-rotation about $S^y$, respectively), many variants of Witten effect  in the 3D SPT bulk and various versions of {\it anomalous} surface quantum Hall effect are defined and systematically investigated.  If charge or spin symmetry reduces to Z$_N$ by considering charge-$N$ or spin-$N$ condensate, instead of the linear response approach, we gauge the charge/spin  symmetry, leading to a {\it dynamical gauge theory} with some remaining global symmetry.  The 3D dynamical gauge theory describes a symmetry-enriched topological phase (SET), \emph{i.e.} a topologically ordered state with global symmetry which admits nontrivial ground state degeneracy depending on spatial manifold topology. For the SPT examples we consider, the corresponding SET states are described by dynamical topological gauge  theory with topological BF term and axionic $\Theta$-term in 3D bulk. And the surface of SET is described by the chiral boson theory with quantum anomaly.  
\end{abstract}
\maketitle

\tableofcontents
 \section{Introduction}
Searching for exotic quantum many-body states is one of the main goals in condensed matter physics.  Physically, all states (or ``phases'' interchangeably) are either gapped states or gapless states, depending on the bulk energy gap between the ground state and first excitation. Recently, considerable attentions have been drawn by both kinds of states.  Some gapless states with strong correlations emerge at quantum critical points without long-lived quasiparticles, in which some new approaches based on  the holographic principle in string theory are introduced into condensed matter physics. On the other hand, the notion of ``quantum entanglement'' becomes an important viewpoint for understanding gapped states\cite{LW0605,KP0604}. By utilizing the well-defined ``local unitary transformation (LU)'' suggested by Chen \emph{et.al.},  all gapped states are classified into two categories: ``short-range entangled states (SRE)'' and ``long-range entangled states (LRE)'' \cite{Chenlong,Chen10,Wenscience}.

For LRE states, there is no canonical LU to connect LRE to a trivial direct product state (a state with zero entanglement range).  The well-known fractional quantum Hall states (FQH)\cite{Tsui82} are a class of LRE states with highly long-range entanglement.   A LRE state generally admits an intrinsic topological order (TO) or ``topological order'' for short.\cite{Wtop,WNtop,Wrig} A TO state is defined by the following features:
ground state degeneracy in a topologically nontrivial closed
manifold,\cite{Wtop,WNtop,Wrig,Wang:2012am} or emergent fermionic/anyonic
excitations,\cite{H8483,ASW8422} or chiral gapless edge
excitations.\cite{H8285,Wedgerev}  If, in addition to a topological order, the
ground state also has a global symmetry, such a state will be referred to as a
``symmetry-enriched topological (SET) phase''.

In contrast, an SRE state can always be adiabatically deformed to a trivial direct product state and thus generically has no TO, which looks quite boring. However, recent rapid progress indicates that some SRE states are quite special for the reason that these SRE states cannot be adiabatically deformed to a direct product state {\it unless} a certain on-site global symmetry group (\emph{i.e.} a global symmetry operation that is a direct product of the operation on each lattice site) is broken, unveiling existence of nontrivial symmetry-protected properties in such SRE states.\cite{Chenlong,Chen10,Wenscience}  This fact leads to the notion of ``symmetry-protected topological phases'' (SPT), which are currently under considerable investigation.  The well-known three-dimensional non-interacting fermionic topological insulator (TI) state is a fermionic SPT in which the surface single massless Dirac fermion is protected by the symmetry group U(1)$\rtimes$Z$^T_2$ where U(1) is charge symmetry related to the fermionic particle number conservation and Z$^T_2$ is time-reversal symmetry.\cite{TI1,TI2,TI3,TI4,TI5,TI6,TI7}  The Haldane phase in an antiferromagnetic Heisenberg spin chain with integer spin and SO(3) spin rotation symmetry is a bosonic SPT.\cite{Haldanephase}  At each edge of the spin chain, a free spin-$1/2$ degree of freedom arises, which is protected by SO(3) symmetry group. In the following, we will only discuss SPT states in boson / spin systems.  In boson/spin systems, interactions are crucial to realize nontrivial SPT states or TO states. Some useful theoretical approaches are recently proposed, such as exactly soluble models\cite{LevinGu,Burnell13,2013arXiv1303.4301C,2013arXiv1305.5851F,2012FrPhy...7..150W,Chen:2012hc, Santos:2013uda, Hung:2012kc}, the fusion category
approach\cite{FNS0428,LW0510}, projective construction\cite{YW12,MM13,YW13}, K-matrix Chern-Simons approach\cite{Kmatrix,Wang:2012am,Lu:2013jqa,Hung:2012nf,Chenggu,Hungwan} and other field theory approaches\cite{VS1258,Xu1,Xu2,Xu3,LW12},  the group cohomology
approach\cite{Chenlong,Chen10,Wenscience}, and modular invariance of edge conformal field theory (CFT)\cite{Ryu13}. 
In the group cohomology classification theory, an SPT state with symmetry group ``G'' in $d$-dimensional spatial lattice is classified by the group cohomology $\mathcal{H}^{d+1}[G,U(1)]$.\cite{Chenlong,Chen10,Wenscience}  The famous Haldane phase is classified by $\mathcal{H}^{2}[SO(3),U(1)]={Z}_2$ indicating that there is only one nontrivial Haldane phase in addition to one trivial phase. In 3D, the ``bosonic topological insulator'' (BTI) with U(1)$\rtimes$Z$_2^T$ are studied by Refs. \onlinecite{VS1258,XS13,MM13,YW13,WS13}. 

 The group cohomology theory\cite{Dijkgraaf:1989pz} provides the elements of the cohomology group to label SPT states, while, it is also interesting to build up the connection between the abstract labels and physical properties (such as electromagnetic response). The following two general approaches are  mainly applied in the community: {\it response theory} and {\it dynamical gauge theory}.  
 
 For U(1) SPT states in 2D which are labeled by $\mathcal{H}^3[U(1),U(1)]=\mathbb{Z}$, an applicable way to understand the ground states is to add an external U(1) gauge field to study the {\it response theory}.  Here the U(1) gauge field is non-dynamical. The resultant response action is a Chern-Simons term with a quantized Hall conductance $\sigma=2k\times\frac{{e^*}^2}{2\pi}$ where $k\in\mathbb{Z}$ and $e^*$ is the fundamental gauge charge carried by bosons.\cite{SLprl,u11,u12,u13,YW12} This integer $k$ is also the integer label defined in the group cohomology classification theory as a one-to-one correspondence.\cite{Wen:2013ue}

On the other hand, a pioneer work by Levin and Gu\cite{LevinGu} leads to a breakthrough. 
They study the 2D Ising paramagnets (SPT states with Z$_2$ symmetry) by fully gauging the Z$_2$ symmetry, which results in a {\it dynamical gauge theory} where the gauge field variables 
become new degrees of freedom.
%
Their gauging procedure indicates that a trivial paramagnet maps to a dynamical Z$_2$ gauge theory (i.e. Z$_2$ toric code), while 
a nontrivial paramagnet maps to a nontrivial dynamical Z$_2$ gauge theory (i.e. Z$_2$ doubled semions). 
This correspondence gives Z$_2$ classification which is consistent with the group cohomology classification $\mathcal{H}^{3}[Z_2,U(1)]=Z_2$, i.e.  one nontrivial SPT state and one trivial state. In a more complicated symmetry group, one can partially gauge a normal subgroup of the global symmetry group of SPT. The resultant dynamical gauge theory may describe an SET state, since some global symmetries remain in the gauged theory.\cite{Hung:2012nf,Mesaros:2012yd,2013PhRvB..87j4406E,Lu:2013jqa}

In this work, we will elaborate the two approaches ``response theory'' and ``dynamical gauge theory'' in many SPT examples which at least have charge symmetry (U(1) or Z$_N$) or spin $S^z$ rotation symmetry (U(1) or Z$_N$)  in two and three dimensions. The symmetry groups in the present work are more relevant to realistic physical systems in condensed matter. 
 More specifically, if both charge and spin symmetry are the simplest continuous U(1), we apply the response theory to study the charge and spin dynamics of the three-dimensional SPT bulk (denoted by $\Sigma^3$), the surface (denoted by $\partial\Sigma^3$) of the $\Sigma^3$ bulk, and the two-dimensional SPT bulk (denoted by $\Sigma^2$). We stress that, instead of looking into microscopic lattice models and utilizing sophisticated mathematical techniques or physical arguments, in the whole discussions of the present work we will attempt to start with the minimal necessary physical input (such as gauge invariance principle, absence of topological order, etc.)  to 
extract the response theory. We will also apply the so-called ``top-down approach'' based on  the ``$K_G$-matrix'' Chern-Simons term (a gauged version of $K$-matrix Chern-Simons   field theory by adding the external electromagnetic field $A^c_\mu$ minimally coupled to charge current and the external ``spin gauge field'' $A^s_\mu$ minimally coupled to spin current).
From this response approach, we will construct many variants of the celebrated Witten effect and also many different versions of quantum Hall effects, depending on the choice of probe fields (external electromagnetic field or external spin gauge field) and the choice of response current (charge current or spin current). By studying the response phenomena case by case, we  emphasize that, although both $\partial\Sigma^3$ and $\Sigma^2$ are two-dimensional,  the response theory on $\partial\Sigma^3$ is realized in an {\it anomalous} fashion in a sense that an extra spatial dimension (deep into the bulk) is required, in sharp contrast to the response theory of SPT defined on $\Sigma^2$ with the same symmetry implementation. The above results will be discussed in Sec. \ref{sec:continuoussymmetry}. Summary of Witten effects and quantum Hall effects are shown in Table \ref{tab:results}. 

If charge or spin symmetry reduces to Z$_N$ by considering charge-$N$ or spin-$N$ condensate, we change our strategy due to the ``Meissner effect'' for charge and spin. Instead of the response approach, we diagnose the SPT states by gauging a normal subgroup (i.e. the charge symmetry and spin $S^z$ rotation symmetry) of the whole symmetry of SPT, resulting in a dynamical gauge theory with both gauge symmetry and global symmetry. In the examples we will consider,  each dynamical gauge theory in $\Sigma^3$ describes an SET state in which ground state degeneracy is nontrivially dependent on the spatial manifold topology. More specifically, the generic form of the dynamical gauge theory in $\Sigma^3$ is a topological gauge theory and consists of two topological terms: topological BF terms and variants of axionic $\Theta$-terms. And its surface ($\partial\Sigma^3$) theory is chiral boson matter field theory that admits quantum anomaly and is meant to cancel the anomaly from $\Sigma^3$ due to the existence of topological BF term. By studying the dynamical gauge theory description of SPT states, we  emphasize that, although both $\partial\Sigma^3$ and $\Sigma^2$ are two-dimensional, the surface dynamical theory on $\partial\Sigma^3$ after promoting the bulk SPT to an SET state can not be realized in a dynamical gauge theory defined on $\Sigma^2$ after promoting the $\Sigma^2$ SPT to a dynamical gauge theory.  The strategy of dynamical gauge theory provides the connection between SPT and SET. Most importantly, if we gauge a normal subgroup and place gauge connection on the lattice links,\cite{LevinGu} we can in principle study the corresponding dynamical gauge theory with boson matter put on lattice in a numerical simulation approach, in order to extract the nature of the underlying SPT state. The above results will be discussed in Sec. \ref{sec:discretesymmetry}. Some key results are collected in Table \ref{tab:discrete} for the reader's convenience.  Sec. \ref{section:conclusion} is devoted to the conclusions of the paper.

\section{Continuous U$_c$(1) charge symmetry and U$_s$(1) spin symmetry}\label{sec:continuoussymmetry}

\subsection{General response theory based on gauge invariance}

\subsubsection{$\Theta$-matrix response theory in $\Sigma^3$}

Let us begin with a three-dimensional bulk of spin-1 and charge-1 boson system where {\it at least} U$_c$(1) and U$_s$(1) are unbroken. Here, the conserved charge corresponding to U$_s$(1) symmetry is the spin density $S^z$. Let us probe the charge and spin dynamics by gauging these two U(1) symmetries rendering two {\it non-dynamical} gauge fields: the ``spin gauge field'' $A^s_\mu$ and the usual electromagnetic gauge field $A^c_\mu$. Note that, the spin gauge field $A^s_\mu$ directly couples to spin density and spin current.

According to the general principle of {\it gauge invariance}, the linear response theory obtained by integrating out bosons should be the following general form in the long-wavelength limit:
\begin{align}
&\mathcal{Z}[A^s_\mu, A^c_\mu]=e^{iS_{\rm top}}\,,S_{\rm top}=\int d^4x\mathcal{L}\,,\nonumber\\
&\mathcal{L}= \frac{\Theta_{IJ}}{8\pi^2}\partial_{\mu}A^{I}_\nu\partial_\lambda A^{J}_\rho\epsilon^{\mu\nu\lambda\rho}\label{thetamatrixaction}
\end{align}
where, the partition function $\mathcal{Z}$ is a functional of the two non-dynamical gauge fields. The indices $I,J=1,2$. $A_\mu^1, A_\mu^2$ denote $A^c_\mu, A^s_\mu$, respectively. The usual Maxwell terms are not written explicitly for the reason that their physical effects are to  renormalize electromagnetic parameters  (dielectric constant and permeability) of the bulk systems. 
 The coefficient $\Theta_{IJ}$ forms a symmetric matrix:
\begin{align}
\Theta=\left( \begin{matrix}\theta_c&\theta_0\\\theta_0&\theta_s\end{matrix} \right)
\end{align}
The Lagrangian  $\mathcal{L}$ can be written in terms of three parts: 
\begin{align}
\mathcal{L}=&\mathcal{L}_c+\mathcal{L}_s+\mathcal{L}_0\,,\nonumber\\
\mathcal{L}_c=&\frac{\theta_c}{8\pi^2}\partial_{\mu}A^{c}_\nu\partial_\lambda A^{c}_\rho\epsilon^{\mu\nu\lambda\rho}\,,\nonumber\\
\mathcal{L}_s=&\frac{\theta_s}{8\pi^2}\partial_{\mu}A^{s}_\nu\partial_\lambda A^{s}_\rho\epsilon^{\mu\nu\lambda\rho}\,,\nonumber\\
\mathcal{L}_0=&\frac{\theta_0}{4\pi^2}\partial_{\mu}A^{c}_\nu\partial_\lambda A^{s}_\rho\epsilon^{\mu\nu\lambda\rho}\,,\label{three}
\end{align}
In the following, the ``electric field'' $\mathbf{E}^c$ and ``magnetic field'' $\mathbf{B}^c$ are constructed from the gauge field $A^c_\mu$ in the usual convention. And, the new notions of ``spin-electric field'' $\mathbf{E}^s$ and ``spin-magnetic field'' $\mathbf{B}^s$ are  specified to the gauge field $A^s_\mu$.


If the spin current and charge current are well-defined on the surface $\partial \Sigma^3$ of the three-dimensional bulk $\Sigma^3$ or on a strictly two-dimensional plane $\Sigma^2$, we define four kinds of quantum Hall effects as shown in Table. \ref{tab:def}. For example,  in the quantum charge-spin Hall effect, the external ``spin gauge field'''s electric field $\mathbf{E}^s$ drives a perpendicular charge current $\mathbf{J}^c$. The Hall conductance is denoted by $\sigma^{cs}$ ($\widetilde{\sigma}^{cs}$) if the Hall effect is on $\Sigma^2$ ($\partial\Sigma^3$).

\begin{table}
 \begin{tabular}[t]{|c|c|c|c|}
 \hline
 \begin{minipage}[t]{1in}Quantum Hall effects\end{minipage}&\begin{minipage}[t]{.8in}Hall condunctance\end{minipage}&\begin{minipage}[t]{.5in}Probe fields\end{minipage}&\begin{minipage}[t]{0.7in}Response current\end{minipage}\\
  \hline
 \begin{minipage}[t]{1in}Quantum charge Hall effect\end{minipage}&\begin{minipage}{.7in}$\sigma^c$, $\widetilde{\sigma}^c$\end{minipage}&$\mathbf{E}^c$&\begin{minipage}[t]{0.7in}$\mathbf{J}^c$\end{minipage}\\
  \hline
 \begin{minipage}[t]{1in}Quantum spin Hall effect\end{minipage}&\begin{minipage}[t]{.7in}$\sigma^s$,  $\widetilde{\sigma}^s$\end{minipage}&$\mathbf{E}^s$&\begin{minipage}[t]{0.7in}$\mathbf{J}^s$\end{minipage}\\
  \hline
 \begin{minipage}[t]{1in}Quantum charge-spin Hall effect\end{minipage}&\begin{minipage}[t]{.7in}$\sigma^{cs}$, $\widetilde{\sigma}^{cs}$\end{minipage}&$\mathbf{E}^s$&\begin{minipage}[t]{0.7in}$\mathbf{J}^c$\end{minipage}\\
 \hline
 \begin{minipage}[t]{1in}Quantum spin-charge Hall effect\end{minipage}&\begin{minipage}[t]{.7in}$\sigma^{sc}$, $\widetilde{\sigma}^{sc}$\end{minipage}&$\mathbf{E}^c$&\begin{minipage}[t]{0.7in}$\mathbf{J}^s$\end{minipage}\\
\hline
 \hline
 \end{tabular}
  \caption{Four different quantum Hall effects}\label{tab:def}
\end{table}
%
%
%

\begin{table*}
 \begin{tabular}[t]{|c|c|c|c||c|}
 \hline
 \begin{minipage}[t]{1in}Axionic Theta angle\end{minipage}&\begin{minipage}[t]{0.9in}Full symmetry group \end{minipage} &\begin{minipage}[t]{1.4in}3D bulk ($\Sigma^3$) response\end{minipage}&\begin{minipage}[t]{1.6in}Surface ($\partial\Sigma^3$) anomalous response\end{minipage}&
 \begin{minipage}[t]{1.4in} 2D plane ($\Sigma^2$) response\end{minipage}\\
 \hline\hline
   \begin{minipage}[t]{1.1in}$\Theta_{11}\equiv\theta_c=2\pi+4\pi k$, charge-1 boson system\end{minipage}&\begin{minipage}[t]{0.5in}U$_c$(1)$\rtimes$Z$^T_2$\end{minipage}&\begin{minipage}[t]{1.4in}charge-Witten effect: $N^c=n^c+N^c_m$\end{minipage}&\begin{minipage}[t]{1.5in}
  Quantum charge Hall effect on Z$^T_2$-broken $\partial\Sigma^3$: $\widetilde{\sigma}^{c}=(1+2k)\frac{1}{2\pi}$\end{minipage}&\begin{minipage}[t]{1.4in}
Quantum charge Hall effect on Z$^T_2$-broken $\Sigma^2$\\ $\sigma^c=2k\frac{1}{2\pi}$\end{minipage}\\
\hline
   \begin{minipage}[t]{1.1in}$\Theta_{22}\equiv\theta_s=2\pi+4\pi k$, spin or boson systems with integer spins \end{minipage}&\begin{minipage}[t]{0.5in}U$_s$(1)$\times$Z$^T_2$\end{minipage}&\begin{minipage}[t]{1.45in}spin-Witten effect: \\$N^s=\sum_i q_in_i^s+N^s_m\sum_{i}q_i^2$\end{minipage}&\begin{minipage}[t]{1.5in}
  Quantum spin Hall effect on Z$^T_2$-broken $\partial\Sigma^3$: $\widetilde{\sigma}^{s}=(1+2k)\frac{1}{2\pi}\sum_i{q_i^2}$\end{minipage}&\begin{minipage}[t]{1.4in}
Quantum spin Hall effect on Z$^T_2$-broken $\Sigma^2$\\ $\sigma^s=2k\frac{1}{2\pi}\sum_i{q_i^2}$\end{minipage}\\
\hline
   \begin{minipage}[t]{1.1in}$\Theta_{12}=\Theta_{21}\equiv\theta_0=\pi+2\pi k$, boson system of charge-1 and spin-1\end{minipage}&\begin{minipage}[t]{1in}U$_c$(1)$ \times $[U$_s$(1)$\rtimes$Z$_2$]\end{minipage}&\begin{minipage}[t]{1.45in}mutual-Witten effect: $N^c=n^c+\frac{1}{2}N^s_m$; $N^s=n^s_{+}-n^s_{-}+\frac{1}{2}N^c_m$\end{minipage}&\begin{minipage}[t]{1.5in}
  Quantum charge-spin / spin-charge Hall effects on Z$_2$-broken $\partial\Sigma^3$: $\widetilde{\sigma}^{cs}=\widetilde{\sigma}^{sc}=(\frac 1 2+k)\frac{1}{2\pi}$\end{minipage}&\begin{minipage}[t]{1.4in}
Quantum charge-spin / spin-charge Hall effects on Z$_2$-broken $\Sigma^2$:\\ $\sigma^{cs}=\sigma^{sc}=k\frac{1}{2\pi}$\end{minipage}\\
 \hline\hline
 \end{tabular}
  \caption{Charge and spin response of spin-1 and charge-1 boson systems (the case of $\Theta_{22}$ is generalized to spin systems of any integer spin $s$ ). $q_i=s,s-1,s-2,\cdots$ and $q_i>0$ where $s$ is total spin defined by $\mathbf{S}^2=s(s+1)$. The full units of $\sigma^c,\sigma^s,\sigma^{cs},\sigma^{sc}$ are: $\frac{e^2}{\hbar}$, $\hbar$, $e$, and $e$, respectively, where, $e$ is elementary electric charge and $\hbar$ is reduced Planck constant. U$_c(1)$ and U$_s(1)$ denote the U(1) symmetry of charge and spin, respectively. Z$_2$ symmetry in ``U$_s(1)\rtimes$Z$_2$'' is the $\pi$-rotation about $S^y$. Z$^T_2$ is time-reversal symmetry. $k\in\mathbb{Z}$, and, ``$\rtimes$'' stands for ``semidirect product''. 
  }\label{tab:results}
\end{table*}

\subsection{U$_c$(1)$\rtimes$Z$^T_2$ in $\Sigma^3$}\label{sec:chargewitten11}

Let us consider the $\Sigma^3$ bulk with U$_c$(1)$\rtimes$Z$_2^T$ symmetry, where, U$_c$(1) and Z$_2^T$ are charge conservation symmetry and time-reversal symmetry, respectively. A bosonic system with this symmetry in $\Sigma^3$ is a {\it bosonic topological insulator} (BTI) which has been recently studied\cite{VS1258,XS13,MM13,YW13,WS13}.  By applying $A^c_\mu$ to probe the topological electromagnetic properties, the resultant response theory gives rise to the topological magneto-electric effect with $\theta_c$ quantized at $2\pi$ mod$(4\pi)$. In the following, we will study the response theory along two approaches. Firstly, based on the response current and the definition of SPT states, we will  derive the quantization of $\theta_c$, charge-Witten effect \cite{Witten79,Franz10,TI7,TI6} in bosonic topological insulator in $\Sigma^3$, and the quantum charge Hall effect on the Z$^T_2$-broken surface $\partial\Sigma^3$ and  Z$^T_2$-broken 2D plane $\Sigma^2$. Partially along the line of the physical arguments of Ref. \onlinecite{MM13},  we shall elaborate the derivation in details in order for the generalization to other symmetry groups (Table \ref{tab:results}) in the remaining parts of Sec. \ref{sec:continuoussymmetry}. Secondly, we shall rederive these results through  the top-down approach by comparing $K_G$-matrix on $\Sigma^2$ and its anomalous realization on $\partial\Sigma^3$ (``$K_G$'' will be defined later). 

\subsubsection{Charge-Witten effect in $\Sigma^3$}

 BTI admits surface charge Hall effect by breaking Z$^T_2$ on the surface. The formation of surface can be viewed as an interface between vacuum and BTI bulk where the derivative of $\theta_c$ forms a two-dimensional domain wall. Let us study the response equation of $A^c$ in the bulk $\theta_c$-term:
\begin{align}
J^c_\mu\equiv\frac{\delta\mathcal{L}_c}{\delta A^c_\mu}=2\times\frac{\theta_c}{8\pi^2}\partial_\nu\partial_\lambda A^c_\rho \epsilon^{\mu\nu\lambda\rho}=\frac{\theta_c}{4\pi^2}\partial_\nu\partial_\lambda A^c_\rho\epsilon^{\mu\nu\lambda\rho}\,,
\end{align}
where, $J^c_\mu$ is (3+1)D response charge current. the prefactor $2$ comes from twice variations with respect to $A^c_\mu$. The zero component $J^c_0$ denotes the response charge density probed by external field $A^c_\mu$
\begin{align}
J^c_0=\frac{\theta_c}{4\pi^2}\nabla\cdot \mathbf{B}^c\,,
\end{align}
where, $\mathbf{B}^c$ is the magnetic field variable. If the gauge field $\mathbf{A}^c$ is smooth everywhere, $\nabla\cdot \mathbf{B}^c=0$ due to absence of magnetic charge. However, if singular configuration is allowed, the divergence may admit singularities in the bulk and its total contribution in the bulk is quantized due to Dirac quantization condition (or more general Schwinger-Zwanziger quantization condition
\cite{Schwinger:1969ib,Zwanziger:1968rs}). In a simplest configuration, let us consider one magnetic monopole which is located at the origin of the three-dimensional space:
 \begin{align}
 \int d^3x \nabla\cdot\mathbf{B}^c=2\pi N_m^c\,,
 \end{align}
where, $N_m^c\in \mathbb{Z}$ is an integer-valued ``magnetic charge'' in $A^c_\mu$ gauge group. Therefore, the corresponding response charge is:
\begin{align}
N^c=\int d^3x J^c_0=\frac{\theta_c}{2\pi}N_m^c\,.
\end{align}
which indicates a nonzero Theta term supports a ``polarization charge cloud'' in the presence of magnetic monopole. A monopole can also trivially attach integer number ($n^c$) of charge-1 bosons in the bulk. Therefore, the whole formula of the so-called {\it charge-Witten effect}\cite{Witten79,Franz10} in Table \ref{tab:results} can be expressed as:
\begin{align} 
N^c=n^c+\frac{\theta_c}{2\pi}N^c_m\,.\label{3dwitten}
\end{align}

\subsubsection{Quantum charge Hall effect on Z$^T_2$-broken $\partial\Sigma^3$ and Z$^T_2$-broken $\Sigma^2$ \label{SecIIB2}}
The bulk $\theta_c$ term $\mathcal{L}_c$ can be written as a surface term:
\begin{align}
\mathcal{L}_{c,\partial\Sigma^3}=\frac{\theta_c}{8\pi^2}A^c_\mu \partial_\nu A^c_\lambda \epsilon^{\mu\nu\lambda}\,.\label{csurface}
\end{align}
which leads to the surface response current:
\begin{align}
{J}^{c,\partial\Sigma^3}_\mu\equiv \frac{\delta \mathcal{L}_{c,\partial\Sigma^3}}{\delta A^c_\mu}=\frac{\theta_c}{4\pi^2}\partial_\nu A^c_\lambda \epsilon^{\mu\nu\lambda}\,.
\end{align}
The surface charge Hall conductance $\sigma^c$ is defined by Ohm's equation $J^{c,\partial\Sigma^3}_x=\widetilde{\sigma}^c E^c_y$ where $E^c_y$ is the electric field along y-direction (assuming that $\partial\Sigma^3$ is parametrized by $x-y$ coordinates):
\begin{align}
\widetilde{\sigma}^c=\frac{\theta_c}{4\pi^2}\label{surfacehall}
\end{align}

To understand the quantization of the surface charge Hall conductance $\widetilde{\sigma}^c$, we need to firstly understand the quantization on $\theta_c$ angle and $\sigma^c$ in a strictly 2D system (i.e. $\Sigma^2$) which is defined as a U$_c$(1) SPT. Let us write down the Chern-Simons term in Z$^T_2$-broken $\Sigma^2$ which describes the response theory of U$_c$(1) SPT on $\Sigma^2$:
\begin{align}
\mathcal{L}_{c,\Sigma^2}=\frac{\sigma^c}{2} A^c_\mu \partial_\nu A^c_\lambda \epsilon^{\mu\nu\lambda}\,.\label{eq:c2}
\end{align}
Upon adiabatically piercing $\Sigma^2$ by $2\pi$ magnetic flux ($\Phi^c=\int d^2  x\nabla\times \mathbf{A}^c=2\pi$), the total response charge $\int d^2x J_0^{c,\Sigma^2}=\sigma^c\int d^2x  \nabla\times \mathbf{A}^c=2\pi\sigma^c$. In SPT states where topological order is trivial by definition ({\it at least} no exotic fractional charge), this pumped charge in the centre of the vortex core must be quantized at integer carried by charge-1 bosons of underlying microscopic model, such that, $2\pi\sigma^c\in\mathbb{Z}$.  This condition is enough for free fermion system. However, for the bosonic system (i.e. U$_c$(1) SPT we are considering), we need further forbid quasiparticles which carry non-bosonic statistics (i.e. fermionic statistics and anyonic statistics) in order to obtain states without topological order. To achieve this goal, one can spatially exchange two vortex cores of $2\pi$ fluxes each of which traps $2\pi\sigma^c$ quasiparticles. The quasiparticles in the first vortex core will perceive a $\pi$ phase as half a magnetic flux of the second vortex core, and, vice versa. The total Aharonov-Bohm phase ``$\Phi_{\rm AB}$'' in the Chern-Simons theory, however, is only {\it half} of the totally accumulated quantum phases: $\Phi_{\rm AB}=\frac{1}{2}\times (2\pi\sigma^c\times\pi+2\pi\sigma^c\times\pi)=2\pi^2\sigma^c$\cite{Wenbook}. In order to forbid non-bosonic statistics, a new condition should be satisfied: $\Phi_{\rm AB}/2\pi\in\mathbb{Z}$. Overall, $2\pi\sigma^c/2\in\mathbb{Z}$, i.e. 
\begin{align}
\sigma^c=2k \frac{1}{2\pi}\,,\label{2dhall}
\end{align}
where, $k\in\mathbb{Z}$. After this preparation, let us move on to the $\theta_c$ angle quantization and its periodicity. Generally, a Theta term is odd under Z$^T_2$ and thus breaks Z$^T_2$ symmetry and thus results in CP-violation in the context of high energy physics \cite{Witten79}, because under Z$^T_2$, $\mathbf{E}^c\rightarrow \mathbf{E}^c\,,\mathbf{B}^c\rightarrow -\mathbf{B}^c$, $\frac{\theta_c}{8\pi^2}\epsilon^{\mu\nu\lambda\rho}\partial_\mu A^c_\nu\partial_\lambda A^c_\rho=\frac{\theta_c}{4\pi^2}\mathbf{E}^c\cdot \mathbf{B}^c\rightarrow -\frac{\theta_c}{4\pi^2}\mathbf{E}^c\cdot \mathbf{B}^c$. However, if $\theta_c$ admits a periodic shift such that $-\theta_c$ can be shifted back to $\theta_c$, the action eventually is time-reversal invariant.  Therefore, the minimal Theta value should be one half of its periodicity (say, ${P}$), and the symmetry group for the bulk is indeed U$_c$(1)$\rtimes$Z$^T_2$ as we defined at the beginning of this section. Physically, the periodicity can be understood as trivially depositing arbitrary copies of $\Sigma^2$ Hall systems onto the surface\cite{VS1258,MM13}. As such, a ${P}$ shift in $\theta_c$ leads to an additional term  in the surface charge Hall conductance formula (\ref{surfacehall}):
\begin{align}
\widetilde{\sigma}^{c \prime}-\widetilde{\sigma}^c=\frac{{P}}{4\pi^2}
\end{align}
which is contributed by deposited $\Sigma^2$ layers which are described by Eq. (\ref{2dhall}). A minimal choice is $\frac{{P}}{4\pi^2}=2\times\frac{1}{2\pi}$, so that ${P}=4\pi$. And the minimal choice of $\theta_c$ is $\frac{{P}}{2}=2\pi$, i.e.:
\begin{align}
\theta_c=2\pi +4\pi k\,.\label{thetacquantize}
\end{align}
where, the integer $k$ is the same $k$ defined in Eq. (\ref{2dhall}).

Substituting Eq. (\ref{thetacquantize}) into Eq. (\ref{surfacehall}) leads to:
\begin{align}
\widetilde{\sigma}^c=(1+2k)\frac{1}{2\pi} \label{3dhall}
\end{align}
The most anomalous phenomenon on $\partial\Sigma^3$ is that the surface quantum charge Hall conductance $\widetilde{\sigma}^c$ admits a $\frac{1}{2\pi}$ value which cannot be realized  in $\Sigma^2$ where $\sigma^c$ is always even integer copies of $1/2\pi$. And, substituting Eq. (\ref{thetacquantize}) into Eq. (\ref{3dwitten}) leads to:
\begin{align}
N^c=n^c+N^c_m\,.
\end{align}
where, $k=0$ is selected for simplicity.  $\theta_c=2\pi$ is topologically distinct from $\theta_c=0$ trivial vacuum once the symmetry group U$_c$(1)$\rtimes$Z$^T_2$ is unbroken. Different choices of $k$ actually correspond to the {\it same phase}\cite{VS1258}.

\subsubsection{Anomalous $K_G$-matrix on $\partial\Sigma^3$ \label{AcKmat}}

In the above discussion, we obtained the charge Hall conductance in $\Sigma^2$ and its anomalous realization in Z$^T_2$-broken $\partial\Sigma^3$ based on the general principle of gauge invariance, response definition, and the definition of SPT states. In those derivations, the microscopic degree of freedoms are not explicitly written in terms of Lagrangian or Hamiltonian. In the following, we shall start with microscopic degrees of freedom, which is called ``top-down approach''.

By definition, the fundamental elements of a given 2D SPT microscopic model are spins or bosonic particles. In the hydrodynamical approach, however, 
the low energy modes dominating the partition function can be effectively replaced by an SET of {\it statistical} one-form U(1) gauge fields $\{a^{I}_\mu\}$ or two-form U(1) gauge fields $\{b^{I}_{\mu\nu}\}$ ($I=1,2,\cdots$) or higher-form gauge fields. In the following, these gauge field variables are dubbed ``intrinsic / statistical  gauge fields'' interchangeably.  Especially in a 2D system, the current operator $J_\mu$ of a point-particle can be expressed as: $J^\mu=\frac{1}{2\pi}\epsilon^{\mu\nu\lambda}\partial_\nu a_\lambda$ which automatically resolves the currrent conservation equation $\partial_\mu J^\mu=0$. 

 What is the generic low energy theory of the SPT state in terms of these dynamical gauge fields? If we only focus on the topological properties of the SPT ground state (or a general ground state with Abelian topological order), the renormalization group flows to an infrared fixed-point field theory in the Chern-Simons form in 2+1D\cite{Zee:2003mt}.
In other words, the low-energy field theory of the microscopic SPT model is effectively described by a generic Chern-Simons theory of $\{a^{I}_\mu\}$ with a $K$-matrix coefficient:
$ \mathcal{L}_{SPT}=\frac{1}{4\pi}K_{IJ}a^{I}_\mu\partial_\nu a^{J}_\lambda\epsilon^{\mu\nu\lambda}$ where, $I,J=1,2,\cdots$.

Thus it is effective to describe the internal microscopic degrees of freedom of SPT states by a generic $K$-matrix  Chern-Simons field theory of $\{a^I_\mu\}$. Based on this low energy field theory, the response theory is straightforward by adding a minimal coupling term $J^\mu A_\mu$ where  $A_\mu$ is an external gauge field and $J^\mu$ is the current operator carrying gauge charge (i.e. the Noether current related to a global symmetry before gauging it) in terms of $\{a^I_\mu\}$.

This top-down approach starts with intrinsic gauge field $a_\mu$ to construct SPT with Lagrangian $\cL_{SPT}(a)$ and then probe it by external field $A_\mu$. The 
Lagrangian becomes $\cL_{SPT+Gauge}(a,A)$. 
By integrating out $\{a^I_\mu\}$ to obtain the low energy physics of external field $A_\mu$, we obtain an effective theory $\cL_{SPT+Gauge}(A)$.
This is exactly what we would like to do in the following - to confirm our previous result (response theory of electromagnetic field $A_\mu^c$ which couples to charge current and ``spin gauge field'' $A_\mu^s$ which couples to spin current) by comparing with the top-down approach (starting from the intrinsic $\{a_\mu\}$ statistical gauge fields). It should be noted that  the field variables $A^s_\mu$ and $A^c_\mu$ are always treated as non-dynamical background fields in the whole Sec. \ref{sec:continuoussymmetry} due to the standard definition of linear response theory.

 The above procedure is known and applied in the literature\cite{LevinGu,Kmatrix,Chenggu,Hung:2012dx,Hung:2012nf,Hungwan,Lu:2013jqa}, however, 
 to be self-contained and make this method more accessible to the research community, in Appendix. \ref{appA},  
 we carry out an explicit derivation for the following steps:
\be 
\cL_{SPT}(a) \to  \cL_{SPT+Gauge}(a,A) \to \cL_{SPT+Gauge}(A)\label{sequence}
\ee

We use $K$-matrix Chern-Simons effective field theory approach to understand this procedure: 
\be
\cL_{SPT}(a)=\frac{1}{4\pi} K_{S,IJ}\epsilon^{\mu\nu\rho} a^I_\mu \partial_\nu a^J_\rho 
\ee
and we end up with 
\be
\cL_{SPT+Gauge}(A)=\frac{1}{4\pi} K_{G,IJ}\epsilon^{\mu\nu\rho} A^I_\mu \partial_\nu A^J_\rho 
\ee
We denote $K_{S}$ and $K_{G}$ as $K$-matrices for $\cL_{SPT}(a)$ and $\cL_{SPT+Gauge}(A)$ respectively. $a_\mu^I$ ($I=1,2,\cdots$) represent a set of intrinsic fields $a_\mu$ in a general case. Each of $A_\mu^I$ $(I=1,2,\cdots)$ represents an external field which couples to the matter current carrying U$_I$(1) gauge charge.  
Our inspiration is from earlier pioneer works.  
In Ref. \onlinecite{Kmatrix}, Lu and Vishwanath focus on $\cL_{SPT}(a)$. 
In Ref. \onlinecite{Chenggu}, Cheng and Gu attempt to apply the braiding statistics in $\cL_{SPT+Gauge}(A)$ to determine 
the classification of SPT. 
Our approach is analogue to the work by Hung and Wan \cite{Hungwan} who had carried out the simplest gauging procedure for Z$_N$ SPT in 2D.
On the other hand, our key focus is to bridge Ref. \onlinecite{Kmatrix} to Refs. \onlinecite{Chenggu,Hungwan} by directly gauging the global symmetry current and then apply to more complicated symmetry groups (in Appendix. \ref{appA}). 

Let us explicitly work out the response theory on 
Z$_2^T$-broken $\partial\Sigma^3$ of 3D bulk with U$_c$(1)$\rtimes$Z$^T_2$, 
and on  Z$_2^T$-broken $\Sigma^2$. 
Here we save detailed derivations to Appendix. \ref{appA} and list down key results directly. 
We firstly study the response theory on $\Sigma^2$ with U$_c$(1) global symmetry. What we start with is the intrinsic SPT's $
K_{S} =\bigl( {\begin{smallmatrix} 
0 &1 \\
1 & 0
\end{smallmatrix}}  \bigl) 
$. By gauging the U(1) global symmetry current coupling to $A^{c}$, we obtain $K_{G} =2p$ and response action 
\be
\cL_{SPT+Gauge}(A^c) =\frac{2p}{4\pi} \epsilon^{\mu\nu\rho} A^{c}_\mu   \partial_\nu A^{c}_\rho
\ee
with $p\in \mathbb{Z}$ labeling the $\mathbb{Z}$ class of the cohomology group $\cH^3(U(1),U(1))=\mathbb{Z}$.
The charge Hall conductance as the quantum feature of the response of this $\cL_{SPT+Gauge}(A^c)$ is
\be
\sigma^c=2p \frac{1}{2\pi}\,,\label{2dhallKmat}
\ee
This result matches exactly as Eq. (\ref{2dhall}). 
On the other hand, the anomalous $K_G$-matrix in $\partial\Sigma^3$ is harder to obtain from the top-down approach -
because the full classification of 3+1D SPT from intrinsic topological field theory is not yet known to be complete (not even fully matching with the group cohomology).
In principle, we should have topological terms (like $g_1 \epsilon^{\mu\nu\rho \tau} \partial_\mu a_\nu \partial_\rho a_\tau+g_2 \epsilon^{\mu\nu\rho \tau} b_{\mu\nu} \partial_\rho a_\tau +$ etc.)  to generate all classes of $\cH^3(\U(1) \rtimes Z^T_2,\U(1))=\text{Z}_2^2$ (or Z$^3_2$ according to other field theory approach\cite{VS1258}). Here we simply adopt the result from the previous section to state the effective $K_{G,\partial\Sigma^3} =2p+\theta_c/2\pi$, which means
\be
\cL_{SPT+Gauge}(A^c) =\frac{2p+\theta_c/2\pi}{4\pi} \epsilon^{\mu\nu\rho} A^{c}_\mu   \partial_\nu A^{c}_\rho
\ee
with surface charge Hall conductance
\be
\tilde{\sigma}^c=(2p+\theta_c/2\pi) \frac{1}{2\pi}\,\label{3dhallKmat}
\ee
Therefore, we have Eq. (\ref{2dhallKmat}) and Eq. (\ref{3dhallKmat}) written in a consistent manner as Eq. (\ref{2dhall}) and Eq. (\ref{3dhall}).  Now, we have two $K_G$ matrices: one is for the surface and one is for the 2D SPT. We dubbed $K_{G,\partial\Sigma^3}$ ``anomalous $K_G$ matrix'' since it has the same symmetry as the 2D SPT but different response theory. The matrices in the present example is nothing but a number; in the example U$_c$(1)$\times$[U$_s$(1)$\rtimes$Z$_2$], the matrices are two-dimensional.

\subsection{U$_s$(1)$\times$Z$^T_2$ in $\Sigma^3$}\label{sec:spinwitten11}

\subsubsection{Quantum spin Hall effect on Z$^T_2$-broken $\partial\Sigma^3$ and Z$^T_2$-broken $\Sigma^2$ \label{SecIIC1}}

Now, let us move on to the bulk $\theta_s$-term which is the response action of three-dimensional SPT protected by U$_s$(1)$\times$Z$^T_2$. Under Z$^T_2$, $\mathbf{E}^s\rightarrow -\mathbf{E}^s,\mathbf{B}^s\rightarrow \mathbf{B}^s$ in contrast to U$_c$(1) gauge fields. And, the gauge charge in U$_s$(1) gauge group will also change sign under Z$^T_2$ due to its nature of pseudo-scalar.  As bosons in a realistic material system carry integer spins, we start with spin-1 boson system in three dimensions with U$_s$(1)$\times$Z$^T_2$. For this simplest case, we can derive all quantities through the same way as in U$_c$(1)$\rtimes$Z$^T_2$. In the following, we shall derive the quantities ($\theta^s, \sigma^s,\widetilde{\sigma}^s$, spin-Witten effect, etc.) through a different way which is helpful for higher integer spin.

Let us embed U$_s$(1) into SU(2) full spin rotation symmetry. Then we shall only focus on the Lie group representations labeled by integer-spin $s$, which forms a complete set of irreducible representations of SO(3) spin rotation symmetry group. It turns out that the physical consideration, i.e.  ``large gauge invariance'' is sufficient to obtain the results.

SU(2) SPT on $\Sigma^2$ admit nonchiral edge states but the symmetry group acts chirally. This edge profile is stable under symmetry-allowed perturbation. For the edge of SU(2) SPT, the {\it projective} Kac-Moody algebra description is $\int dx  \frac{2\pi}{3}(v_L:\mathbf{J}_L\cdot \mathbf{J}_L:+v_R:\mathbf{J}_R\cdot \mathbf{J}_R:)\,$ where $v_L$ / $v_R$ is velocity, and, the left-mover $\mathbf{J}_L$ is SU(2) doublet and the right-mover $\mathbf{J}_R$ is SU(2) singlet.\cite{YW12,LW12}  After gauging SU(2), the resultant gauge theory after integrating out all matter field fluctuations must be a generic Chern-Simons form in 2+1 D space-time, i.e. (coupling constant $g=1$)
\begin{align}
\mathcal{L}_{s,\Sigma^2}=\frac{k}{4\pi}\epsilon^{\mu\nu\lambda} \text{Tr}\left[A_{\mu}\partial_{\nu}A_\lambda+\frac{2}{3}A_\mu A_\nu A_\lambda\right]
\end{align}
where, $k$ is quantized at integer since it is the winding number labeling the nontrivial homotopic mapping $\pi_3$(SU(2))=$\mathbb{Z}$.  $A_\mu$ is the matrix-valued gauge vector defined as $A_\mu=T^aA^a_\mu$ ($a=1,2,3$), where, $T^a$ are the generators of SU(2) in a given representation (labeled by total spin $s$), and, $A^a_\mu$ are  the real-number valued gauge potentials. The field strength $F_{\mu\nu}=\partial_\mu A_\nu-\partial_\nu A_\mu-i[A_\mu,A_\nu]$.  The Lie algebra structure constant $f^{abc}=\epsilon^{abc}$, i.e. $[T^a, T^b]=i\epsilon^{abc}T^c$. And most importantly, the trace $\text{Tr}[T^aT^b]$ depends on the choice of representation, i.e. the total spin-$s$:
\begin{align}
\text{Tr} [T^aT^b]=\frac{1}{3}s(s+1)(2s+1)\delta_{ab}\,,
\end{align}
Such a normalization condition as well as the structure constant leads to the fact that the generators $T^a$ are {\it precisely} identical to spin operators $S^a$ of spin-$s$ ($(T^1)^2=(T^2)^2=(T^3)^2=\frac{1}{3} s(s+1)\mathbb{I}$ with $\mathbb{I}$ the $2s+1$-dimensional identity matrix). As a result, the gauge field $A^a_\mu$ {\it precisely} couples to spin current of spin-$s$ along $a$-spin direction in a correct unit.

In order to probe the quantum spin Hall effect of the bosonic system, it is sufficient to merely consider a U(1) subgroup of SU(2) in a given representation. For example, let us study the Hall current generated by $A^3_\mu$ (which is identical to $A^s_\mu$):
\begin{align}
J^{s,\Sigma^2}_\mu\equiv \frac{\delta \mathcal{L}_{s,\Sigma^2}}{\delta A^s_\mu}
\end{align}
For convenience, let us express Chern-Simons action explicitly in terms of $A^a_\mu$:
\begin{align}
\mathcal{L}_{s,\Sigma^2}=&\frac{k}{4\pi}\epsilon^{\mu\nu\lambda} \text{Tr}\left[A_{\mu}\partial_{\nu}A_\lambda+\frac{2}{3}A_\mu A_\nu A_\lambda\right]\nonumber\\
=&\frac{k}{4\pi} \epsilon^{\mu\nu\lambda} \text{Tr}\left[ A^a_{\mu}\partial_\nu A^b_\lambda T^aT^b+\frac{2}{3}A_\mu A_\nu A_\lambda\right] 
\end{align}
 Since we only consider the response action of $A^3_\mu$($\equiv A^s_\mu$). Let us drop all terms containing $A^1_\mu, A^2_\mu$:
\begin{align}
\mathcal{L}_{s,\Sigma^2}=\frac{k}{4\pi}  \epsilon^{\mu\nu\lambda} A^s_\mu\partial_\nu A^s_{\lambda} \text{Tr}[T^3T^3]
\end{align}
Therefore, the spin Hall current is given by:
\begin{align}
J^{s,\Sigma^2}_\mu=\frac{k}{2\pi} \text{Tr}[T^3T^3]\epsilon^{\mu\nu\lambda}\partial_\nu A^s_\lambda
\end{align}
The spin Hall conductance $\sigma^s$ is readily given by:
\begin{align}
\sigma^s=&\frac{k}{2\pi}\frac{1}{3}s(s+1)(2s+1)\nonumber\\
=&2k\times \left(\frac{1}{6}s(s+1)(2s+1)\right)\frac{1}{2\pi}=2k\frac{1}{2\pi}\sum_i q_i^2\label{2dhall1}
\end{align}
where, $q_i=s,s-1,s-2,...$ and $q_i>0$.
In the presence of $\sum_i q^2_i$, we find that the $ \sum_i q^2_i\frac{1}{2\pi}$ is nothing but the unit of spin Hall conductance when ``spin charge $q_i$'' is generically other number than 1. For a spin-$s$ boson system, there are $\mathcal{N}_f=2s+1$ flavors of bosons. Each of them contributes the same even number $2k$ in total spin Hall response. We stress that in the present approach the even integer ``$2k$'' arises naturally as a result of large gauge invariance. As a simple check, when $s=1/2$, $\sigma^s=\frac{k}{4\pi}=2k\times \left(\frac{1}{2}\right)^2\frac{1}{2\pi}$. When $s=1$, $\sigma^s=\frac{k}{\pi}=2k\times (1)^2\frac{1}{2\pi}$. When $s=3/2$, $\sigma^s=\frac{5k}{2\pi}=2k\times (\left(\frac{3}{2}\right)^2+\left(\frac{1}{2}\right)^2)\frac{1}{2\pi}$. The first two results ($s=1/2\,,\,1$) were derived by Liu and Wen \cite{LW12} through the principal chiral non-linear sigma models in which $s=1/2$ and $s=1$ are discussed in SU(2) and SO(3) SPT states respectively. Actually, for the response theory itself, it turns out the different results between SU(2) and SO(3) actually depends on the trace normalization of generators (Tr$[T^aT^b]$). Indeed,   the result that $\sigma^s$ of SO(3) is four times of $\sigma^s$ of SU(2) originate from trace normalization (Tr$[T^aT^b]$=$\frac{1}{2}\delta_{ab}$, if $s$=1/2; $[T^aT^b]$=$2\delta_{ab}$, if $s$=1).
We comment that a more mathematical exposition on this $1/4$ factor quantization difference between SU(2) and SO(3) is found in Sec.4 of Ref.\onlinecite{Dijkgraaf:1989pz} and Sec.2 of Ref.\onlinecite{Moore:1989yh}. Here we use a rather physical language instead, comparing to the more mathematically oriented formalism in Ref.\onlinecite{Dijkgraaf:1989pz,Moore:1989yh}.

Let us focus on integer spin-$s$.  In the $s=1$ case, all results can also be derived if the method in the derivation of $\sigma^c$ is adopted. In a generic $s$ case, the even integer $2k$ directly leads to the absence of topological order in $\Sigma^2$ in a sense that:
\begin{itemize}
\item  All spin excitations carry integer-valued spin-$S^z$ quantum number, no fractional quantum number
\item All spin excitations are bosonic with the total Aharonov-Bohm phase $\Phi_{\rm AB}=2\pi k$ accumulated by spatially exchanging two vortex cores. The spin angular momentum doesn't contribute fermionic sign since $s$ is integer.
\end{itemize}

Next let us consider an SPT with SU(2)$\times$Z$^T_2$ in $\Sigma^3$ in the integer spin-$s$ representation and then only consider the U$_s$(1) subgroup through which we will obtain the surface Chern-Simons term $\mathcal{L}_{s,\partial\Sigma^3}$. With Z$^T_2$ symmetry, the bulk should be a non-Abelian Theta term. Generically, the bulk non-Abelian Theta action is $\overline{\Theta}\times \mathcal{P}$ where $\mathcal{P}$ is the integer-valued Pontryagin index (the bar in the symbol $\overline{\Theta}$ is to distinguish $\overline{\Theta}$ and $\Theta$-matrix defined in Eq. (\ref{thetamatrixaction})),
\begin{align}
S=\overline{\Theta}\times \mathcal{P}=\int d^4 x \overline{\Theta} \frac{1}{16\pi^2}\text{Tr} [F_{\mu\nu}\widetilde{F}_{\mu\nu}]\label{thetaterm}
\end{align}
where, the dual tensor $\widetilde{F}_{\mu\nu}=\frac{1}{2}\epsilon^{\mu\nu\lambda\rho}F_{\lambda\rho}$. $F_{\mu\nu}$ can be further written as $F_{\mu\nu}=T^aF^a_{\mu\nu}$ $
F^a_{\mu\nu}=\partial_\mu A^a_\nu-\partial_\nu A^a_\mu+\epsilon^{abc}A^b_\mu A^c_\nu $. 
 Therefore, the Lagrangian form of Eq. (\ref{thetaterm}) is reformulated to
\begin{align}
\mathcal{L}_{s,\Sigma^3}=\frac{ 2\overline{\Theta}\sum_i q_i^2}{16\pi^2}F^a_{\mu\nu}\widetilde{F}^a_{\mu\nu}\,.\label{thetaterm1}
\end{align} 
The dual tensor for each component is defined as $\widetilde{F}^a_{\mu\nu}=\frac{1}{2}\epsilon^{\mu\nu\lambda\rho}F^a_{\lambda\rho}$.  In Eq. (\ref{thetaterm1}), despite the summation over the three spin directions is independent, the three spin directions actually couple to each other due to the last {\it nonlinear} term in $F_{\mu\nu}$. 
 As a matter of fact, the $\overline{\Theta}$ term is a total-derivative term which doesn't generate topological bulk response if $A_\mu$ is smooth without singularities. Let us consider a material with a surface. Suppose that the surface breaks Z$^T_2$ explicitly or spontaneously, the surface has spin rotational symmetry and the response theory is again a non-Abelian Chern-Simons theory. Mathematically, starting from Eq. (\ref{thetaterm}), we find its surface action:
\begin{align}
S=\frac{2\overline{\Theta}}{16\pi^2}\epsilon^{\mu\nu\lambda}\text{Tr}[A_\mu \partial_\nu A_\lambda+\frac{2}{3}A_\mu A_\nu A_\lambda]\,, 
\end{align}
Completion of trace operation leads to:
\begin{align}
S=\frac{2\overline{\Theta}}{8\pi^2}\sum_i q_i^2 \epsilon^{\mu\nu \lambda}A_\mu^a\partial_\nu A^a_\lambda +\cdots
\end{align}
where, ``$\cdots$'' are nonlinear response terms. Let us only consider the $A^3_\mu\equiv A^s_\mu$ response theory, so that, $2\overline{\Theta}\equiv \theta_s$ leading to the surface spin Hall conductance:
\begin{align}
\widetilde{\sigma}^s=\frac{\theta_s}{2\pi}\sum_i q_i^2\frac{1}{2\pi}\label{conductance1}
\end{align}
To derive the periodicity and minimal value of $\theta_s$ for nontrivial phase with U$_s$(1)$\times$Z$^T_2$ in $\Sigma^3$, we apply the same strategy in the derivation of $\theta_c$. The minimal theta value should be one half of its periodicity (say, ${P}$), and the symmetry group for the bulk is U$_s$(1)$\times$Z$^T_2$. Physically, the periodicity can be understood as trivially depositing arbitrary copies of $\Sigma^2$ Hall systems onto the surface. As such, a ${P}$ shift in $\theta_s$ leads to an additional term  in the surface spin Hall conductance formula (\ref{conductance1}):
\begin{align}
\widetilde{\sigma}^{s \prime}-\widetilde{\sigma}^s=\frac{{P}}{4\pi^2}\sum_i q_i^2
\end{align}
which is contributed by deposited $\Sigma^2$ layers which are described by Eq. (\ref{2dhall1}). A minimal choice is $\frac{{P}}{4\pi^2}=2\times\frac{1}{2\pi}$, so that ${P}=4\pi$. And the minimal choice of $\theta_s$ is $\frac{{P}}{2}=2\pi$, i.e.:
\begin{align}
\theta_s=2\pi +4\pi k\,.\label{thetacquantize1}
\end{align}
where, the integer $k$ is the same $k$ defined in Eq. (\ref{2dhall1}). Substituting Eq. (\ref{thetacquantize1}) into Eq. (\ref{conductance1}) leads to:
\begin{align}
\widetilde{\sigma}^s=(1+2k)\frac{1}{2\pi}\sum_i q_i^2 \label{3dhalls}
\end{align}

\subsubsection{Spin-Witten effect in $\Sigma^3$}

In order to derive the so-called {\it spin-Witten effect} in Table \ref{tab:results} for generic spin $s$, by noting that $2\bar{\Theta}=\theta_s$, let us drop all terms irrelevant to $A^z_\mu$ (i.e. $A^s_\mu$) in Eq. (\ref{thetaterm1}), resulting in:
\begin{align}
\mathcal{L}_{s}=\frac{\theta_s\sum_{i}q^2_i}{8\pi^2}\partial_\mu A^{s}_\nu\partial_\lambda A^s_\rho\epsilon^{\mu\nu\lambda\rho}
\end{align}
Once $s=1$, we obtain $\sum_i q^2_i=1$ and above Lagrangian $\mathcal{L}_s$ is back to the original version of the bulk $\theta_s$-term defined in Eq. (\ref{three}). Here, we would like to consider a generic $s$ which leads to the following response equation for $A^s_\mu$ in $\Sigma^3$:
\begin{align}
J^s_\mu\equiv \frac{\delta \mathcal{L}_s}{\delta A^s_\mu}=\frac{\theta_s\sum_{i}q^2_i}{4\pi^2}\epsilon^{\mu\nu\lambda\rho}\partial_\nu \partial_\lambda A^s_\rho
\end{align}
where, $J^s_\mu$ is (3+1)D response spin current. The zero component $J^s_0$ denotes the response charge density probed by external spin gauge field $A^s_\mu$
\begin{align}
J^s_0=\frac{\theta_s\sum_{i}q^2_i}{4\pi^2}\nabla\cdot \mathbf{B}^s
\end{align}
where, $\mathbf{B}^s$ is the spin-magnetic field variable. If the gauge field $\mathbf{A}^s$ is smooth everywhere, $\nabla\cdot \mathbf{B}^s=0$ due to absence of magnetic charge. However, if singular configuration is allowed, the divergence may admit singularities in the bulk and its total contribution in the bulk is quantized due to Dirac quantization condition (or more general Schwinger-Zwanziger quantization condition). Let us consider one magnetic monopole (of $A^s_\mu$ gauge group) located at the origin of the three-dimensional space:
\begin{align}
\int d^3x \nabla\cdot\mathbf{B}^s=2\pi N_m^s\,,
\end{align}
where $N_m^s\in \mathbb{Z}$ is an integer-valued ``magnetic charge'' of $A^s_\mu$ gauge group. Therefore, the corresponding response total spin is:
\begin{align}
N^s=\int d^3x J^s_0=\frac{\theta_s\sum_{i}q^2_i}{2\pi}N_m^s\,.
\end{align} 
which indicates that a nonzero Theta term supports a ``{\it polarization spin cloud}'' in the presence of magnetic monopole of $A^s_\mu$ gauge group. A monopole of $A^s_\mu$ gauge group can also trivially attach integer number ($n^s_i\in\mathbb{Z}$)of bosons with $S^z=q_i$ in the bulk (as $q_i>0$, a negative $n_i^s$ implies that $|n^s_i|$ bosons in the spin state ``$S^z=-q_i$''). Therefore, the whole formula of the so-called {\it spin-Witten effect} can be expressed as:
\begin{align} 
N^s=\sum_i n^s_i q_i+\frac{\theta_s}{2\pi}N^s_m\sum_i q^2_i\,.\label{3dspinwitten}
\end{align}
Substituting (\ref{thetacquantize1}) into (\ref{3dspinwitten}) leads to:
\begin{align} 
N^s=\sum_i n^s_i q_i+N^s_m\sum_i q^2_i\,.\label{3dspinwitten1}
\end{align}
where, $k=0$ is selected for simplicity. Different choices of $k$ actually correspond to the {\it same phase}.

\subsubsection{Anomalous $K_G$-matrix on $\partial\Sigma^3$\label{AsKmat}}

Similar to Sec \ref{AcKmat},
let us use the top-down approach explicitly working out the external field theory 
on $\partial\Sigma^3$ of U$_s$(1)$\times$Z$^T_2$ global symmetry with Z$^T_2$ symmetry broken, 
and the external field theory 
on $\Sigma^2$ of U$_s$(1) global symmetry.
Here we save detailed derivations to Appendix. \ref{appA} and list down key results directly. 
We firstly study on $\Sigma^2$ with U$_s$(1) global symmetry, what we start with is the intrinsic SPT's $
K_{S} =\bigl( {\begin{smallmatrix} 
0 &1 \\
1 & 0
\end{smallmatrix}}  \bigl) 
$ and gauging the U(1) global symmetry current coupling to $A^{s}$, we obtain $K_{G} =2p$. 
The only difference from Sec. \ref{AcKmat} is that the spin gauge field contributes a factor of $\sum_i q_i^2$, which simply sums over all the spin contribution,
$q_i=s,s-1,s-2,...$ and $q_i>0$, 
\be
\cL_{SPT+Gauge}(A^s) =\frac{2p}{4\pi} \epsilon^{\mu\nu\rho} A^{s}_\mu   \partial_\nu A^{s}_\rho \sum_i q_i^2
\ee
with $p\in \mathbb{Z}$ labeling the $\mathbb{Z}$ class of the cohomology group $\cH^3(\U(1),\U(1))=\mathbb{Z}$.
The Hall conductance as the response of this $\cL_{SPT+Gauge}(A^c)$ is
\be
\sigma^s=2p \frac{1}{2\pi} \sum_i q_i^2 \,,\label{2dShallKmat}  
\ee
This result matches exactly as Eq. (\ref{2dhall1}). 
On the other hand, for the anomalous $K_G$-matrix in $\partial\Sigma^3$,
so far the intrinsic topological field theory of $\Sigma^3$ in 3+1D is not yet known to be completed,  
we simply adopt the result from the previous section to state the effective $K_{G,\partial\Sigma^3} =2p+\theta_s/2\pi$, which means
\be
\cL_{SPT+Gauge}(A^s) =\frac{2p+\theta_s/2\pi}{4\pi} \epsilon^{\mu\nu\rho} A^{c}_\mu   \partial_\nu A^{c}_\rho \sum_i q_i^2
\ee
with Hall conductance
\be
\tilde{\sigma}^s=(2p+\theta_s/2\pi) \frac{1}{2\pi}\sum_i q_i^2 \,\label{3dShallKmat}
\ee
Therefore, we have Eq. (\ref{2dShallKmat}) and Eq. (\ref{3dShallKmat}) written in a consistent manner as  Eq. (\ref{2dhall1}) and Eq. (\ref{3dhalls}). $K_{G,\partial{\Sigma}^3}$ is the anomalous $K_G$ matrix of the present surface state.

\subsection{U$_c$(1)$\times$[U$_s$(1)$\rtimes$Z$_2$] in $\Sigma^3$}\label{sec:mutualresponse}

\subsubsection{Quantum charge-spin / spin-charge Hall effect on Z$_2$-broken $\partial\Sigma^3$ and Z$_2$-broken $\Sigma^2$  \label{SecIID1} }

In the above discussions, we have studied the bulk $\theta_c$ and $\theta_s$ terms each of which is constructed by one kind of gauge field. In the following, we shall consider the bulk $\theta_0$-term $\mathcal{L}_0$ in which $A^s_\mu$ and $A^c_\mu$ are both involved. The {\it minimal} symmetry requirement of this term is U$_c$(1)$\times[$U$_s$(1)$\rtimes$Z$_2$], where Z$_2$ can be viewed as a $\pi$-rotation about spin $S^y$. The additional Z$_2$ is required by the following observation. Under Z$_2$ operation, $A_\mu^s\rightarrow -A^s_\mu$, and thus $\mathbf{E}^s\rightarrow -\mathbf{E}^s\,,\mathbf{B}^s\rightarrow -\mathbf{B}^s\,,\mathbf{E}^c\rightarrow \mathbf{E}^c\,,\mathbf{B}^c\rightarrow \mathbf{B}^c$, such that, $\mathcal{L}_0=\frac{\theta_0}{4\pi^2}\partial_{\mu}A^{c}_\nu\partial_\lambda A^{s}_\rho\epsilon^{\mu\nu\lambda\rho}=\frac{\theta_0}{4\pi^2} (\mathbf{E}^c \cdot\mathbf{B}^s+ \mathbf{B}^c\cdot\mathbf{E}^s )\rightarrow -\mathcal{L}_0$. However, if a periodicity in $\theta_0$ is allowed, $-\theta_0$ will be shifted back to $\theta_0$ leading to the invariance of $\mathcal{L}_0$ under Z$_2$ spin rotation. Due to the existence of the periodicity, we expect that Z$_2$ symmetry plays a similar role in determining quantization conditions of $\theta_0$ and related Hall effects, in comparison with the role of Z$^T_2$ in the bulk  $\theta_c$ and $\theta_s$ terms. Note that, a Theta term with only one kind of gauge field (such as $\theta_c$ and $\theta_s$-terms) in a three-dimensional insulator can be formally viewed as an expectation value of divergence of chiral current in the context of the Adler-Bardeen-Bell-Jackiw anomaly\cite{anomaly1,anomaly2} in a chiral gapless system.

For simplicity, we restrict our attention on a spin-1 and charge-1 boson system in $\Sigma^3$ with symmetry U$_c$(1)$\times[$U$_s$(1)$\rtimes$Z$_2$]. Since $\mathcal{L}_0$ is a total derivative term, the bulk response is trivial unless at least one of the gauge field configuration admits singularities. Let us firstly consider a Z$_2$-broken surface $\partial\Sigma^3$. The Lagrangian $\mathcal{L}_0$ can be written as a {\it surface   mutual Chern-Simons} term:
\begin{align}
\mathcal{L}_{0,\partial\Sigma^3}=\frac{\theta_0}{4\pi^2}A^s_\mu\partial_\nu A^c_\lambda\epsilon^{\mu\nu\lambda}\label{0surface}
\end{align}
Recently, mutual Chern-Simons term with dynamical or non-dynamical gauge fields has been studied in other contexts\cite{mcs1,mcs2,mcs3,mcs4,mcs5,BW}.  The surface mutual Chern-Simons term (\ref{0surface}) leads to the following two different surface response currents:
\begin{align}
&J^{s,\partial \Sigma^3}_\mu\equiv \frac{\delta \mathcal{L}_{0,\partial\Sigma^3}}{\delta A^s_\mu}=\frac{\theta_0}{4\pi^2}\partial_\nu A^c_\lambda\epsilon^{\mu\nu\lambda}\,,\label{current1}\\
&J^{c,\partial \Sigma^3}_\mu\equiv \frac{\delta \mathcal{L}_{0,\partial\Sigma^3}}{\delta A^c_\mu}=\frac{\theta_0}{4\pi^2}\partial_\nu A^s_\lambda\epsilon^{\mu\nu\lambda}\,.\label{current2}
\end{align}
Here, the surface spin current $J^{s,\partial\Sigma^3}_\mu$ is induced by applying external electromagnetic gauge field $A^c_\mu$ rendering the quantum spin-charge Hall effect; while the surface charge current $J^{c,\partial\Sigma^3}_\mu$ is induced by applying external spin gauge field $A^s_\mu$ rendering the quantum charge-spin Hall effect. The corresponding surface spin-charge / charge-spin Hall conductance formulas are:
\begin{align}
\widetilde{\sigma}^{sc}=\widetilde{\sigma}^{cs}=\frac{\theta_0}{4\pi^2}\,\label{surface0}
\end{align}
We note that the two quantum Hall effects share the same Hall conductance and the same unit ``$e$''. 
To understand the quantization of the surface spin-charge and charge-spin Hall conductance $\widetilde{\sigma}^{sc},\widetilde{\sigma}^{cs}$, we need to firstly understand the quantization on $\theta_0$ angle and $\sigma^{sc},\sigma^{cs}$ in strictly 2D systems (i.e. $\Sigma^2$). The $\Sigma^2$ system will be used to determine the periodicity and the minimal value of $\theta_0$ by depositing it onto the Z$_2$-broken surface $\partial\Sigma^3$, so that the minimal symmetry requirement on $\Sigma^2$ is U$_c$(1)$\times$U$_s$(1). 
Let us write down a generic topological response theory in $\Sigma^2$:
\begin{align}
\mathcal{L}_{0,\Sigma^2}=\frac{1}{2}(A^c_\mu, A^s_\mu) \left( \begin{matrix}\sigma^c&\sigma^{cs}\\\sigma^{sc}&\sigma^s\end{matrix} \right)\partial_\nu\left( \begin{matrix}A^c_\lambda\\A^s_\lambda\end{matrix} \right)  \epsilon^{\mu\nu\lambda}\,.
\end{align}
which leads to three independent Chern-Simons terms. Therefore, in general, we need three quantities $\sigma^c,\sigma^s,\sigma^{cs}$ ($\sigma^{cs}=\sigma^{sc}$) to label the quantum Hall states in $\Sigma^2$. But the state we are considering will be applied to be deposited onto the surface $\partial\Sigma^3$ where only mutual Chern-Simons term exists as shown in Eq. (\ref{0surface}). In other words, we shall consider the $\Sigma^2$ system with $\sigma^s=0$ and $\sigma^c=0$, rendering a mutual Chern-Simons term for $\Sigma^2$:
\begin{align}
\mathcal{L}_{0,\Sigma^2}=\sigma^{sc} A^s_\mu \partial_\nu A^c_\lambda   \epsilon^{\mu\nu\lambda}\,. \label{u1u1z2CS}
\end{align}
Thus the response currents are:
\begin{align}
&J^{s, \Sigma^2}_\mu\equiv \frac{\delta \mathcal{L}_{0,\Sigma^2}}{\delta A^s_\mu}=\sigma^{sc}\partial_\nu A^c_\lambda\epsilon^{\mu\nu\lambda}\,,\label{current11}\\
&J^{c, \Sigma^2}_\mu\equiv \frac{\delta \mathcal{L}_{0,\Sigma^2}}{\delta A^c_\mu}=\sigma^{cs}\partial_\nu A^s_\lambda\epsilon^{\mu\nu\lambda}\,.\label{current22}
\end{align}

Upon adiabatically piercing $\Sigma^2$ by $2\pi$ magnetic flux ($\Phi^c=\int d^2  x\nabla\times \mathbf{A}^c=2\pi$), the total spin accumulation $\int d^2x J_0^{s,\Sigma^2}=\sigma^{sc}\int d^2x  \nabla\times \mathbf{A}^c=2\pi\sigma^{sc}$. In SPT states where topological order is trivial by definition, this pumped spin in the centre of the vortex core must be quantized at integer since the fundamental spin is carried by spin-1 bosons (for example, a spin-1/2 quasiparticle is not allowed), such that, $2\pi\sigma^{sc}\in\mathbb{Z}$.   On the other hand, let us consider the condition under which all the quasiparticles are bosonic in order to forbid topological order. To achieve this goal, one can spatially exchange two vortex cores of $2\pi$ fluxes each of which traps $2\pi\sigma^{sc}$ quasiparticles with pure spins and neutral charge. The quasiparticles in the first vortex core will perceive a $\pi$ phase as half a magnetic flux of the second vortex core, and, vice versa. Unlike the Chern-Simons theory, the total Aharonov-Bohm phase ``$\Phi_{\rm AB}$'' in the mutual Chern-Simons theory is the totally  accumulated quantum phases: $\Phi_{\rm AB}= (2\pi\sigma^{sc}\times\pi+2\pi\sigma^{sc}\times\pi)=4\pi^2\sigma^{sc}$. In order to forbid non-bosonic statistics, $\Phi_{\rm AB}/2\pi\in\mathbb{Z}$. Overall, combining the ``absence of fractional spin'' and ``absence of non-bosonic statistics'' leads to, still, $2\pi\sigma^{sc}\in\mathbb{Z}$, i.e. 
\begin{align}
\sigma^{sc}=\sigma^{cs}=k \frac{1}{2\pi}\,,\label{2dhall0}
\end{align}
where, $k\in\mathbb{Z}$ and unit is the fundamental electric charge ``$e$''.

After this preparation, let us move on to the $\theta_0$ angle quantization and its periodicity $P$.  As mentioned, the periodicity can be understood as trivially depositing arbitrary copies of $\Sigma^2$ Hall systems onto the surface $\partial\Sigma^3$. As such, a $P$ shift in $\theta_0$ leads to an additional term in the surface spin-charge /charge-spin Hall conductance formula (\ref{surface0}):
\begin{align}
\widetilde{\sigma}^{sc \prime}-\widetilde{\sigma}^{sc}=\frac{{P}}{4\pi^2}
\end{align}
which is contributed by deposited $\Sigma^2$ layers which are described by Eq. (\ref{2dhall0}). A minimal choice is $\frac{{P}}{4\pi^2}=1\times\frac{1}{2\pi}$, so that ${P}=2\pi$. And the minimal choice of $\theta_0$ is $\frac{{P}}{2}=\pi$, i.e.:
\begin{align}
\theta_0=\pi +2\pi k\,.\label{theta01}
\end{align}
where, the integer $k$ is the same $k$ defined in Eq. (\ref{2dhall0}).

Substituting Eq. (\ref{theta01}) into Eq. (\ref{surface0}) leads to:
\begin{align}
\widetilde{\sigma}^{sc}=\widetilde{\sigma}^{cs}=(\frac{1}{2}+k)\frac{1}{2\pi}\,. \label{3dhall0}
\end{align}
The most anomalous phenomenon in the surface charge Hall effect is that the $\widetilde{\sigma}^{sc}$ and $\widetilde{\sigma}^{cs}$ admit a $\frac{1}{4\pi}$ value which cannot be realized  in $\Sigma^2$ where $\sigma^{sc}$ and $\sigma^{cs}$  are always integer copies of $1/2\pi$.

\subsubsection{Model construction on $\Sigma^2$}

 The quantum charge-spin / spin-charge Hall effects can be modeled as a similiar two-component boson model proposed by Senthil and Levin \cite{SLprl} but with slight modification as follows. The first-component bosons are charge neutral but carry spin-1, while, the second-component bosons are spinless but carry electric charge-1. Then an external ``spin-magnetic field'' $\mathbf{B}^s$ and an external real ``magnetic field'' $\mathbf{B}^c$ are applied to the first-component and second-component bosons, respectively, each of which forms a bosonic Landau level with filling $\nu=1$.  
It implies that the gauge fields $\mathbf{A}^s$ and $\mathbf{A}^c$ play the roles of $\mathbf{A}_{1}$ and $\mathbf{A}_{2}$ defined in Senthil-Levin paper, respectively. 
 Therefore, by using the
 flux-attachment Chern-Simons theory, one can realize a many-body state which has quantum spin-charge / charge-spin Hall effects with $\sigma^{sc}=\sigma^{cs}=\frac{1}{2\pi}$.

\subsubsection{Mutual-Witten effect in $\Sigma^3$}

In order to derive the so-called {\it mutual-Witten effect} in Table \ref{tab:results}, let us write down the response equation in the bulk $\Sigma^3$:
\begin{align}
J^s_\mu\equiv \frac{\delta \mathcal{L}_0}{\delta A^s_\mu}=\frac{\theta_0}{4\pi^2}\epsilon^{\mu\nu\lambda\rho}\partial_\nu \partial_\lambda A^c_\rho\,,\\
J^c_\mu\equiv \frac{\delta \mathcal{L}_0}{\delta A^c_\mu}=\frac{\theta_0}{4\pi^2}\epsilon^{\mu\nu\lambda\rho}\partial_\nu \partial_\lambda A^s_\rho\,,
\end{align}
where, $J^s_\mu$ and $J^c_\mu$ are (3+1)D response spin and charge currents, respectively. The zero components $J^s_0$ and $J^c_0$ denote the response spin and charge density probed by external spin gauge field $A^s_\mu$ and external electromagnetic field $A^c_\mu$:
\begin{align}
J^s_0=\frac{\theta_0}{4\pi^2}\nabla\cdot \mathbf{B}^c\,,\\
J^c_0=\frac{\theta_0}{4\pi^2}\nabla\cdot \mathbf{B}^s\,,
\end{align}
If the gauge fields $\mathbf{A}^c$ and $\mathbf{A}^s$ are smooth everywhere, $\nabla\cdot \mathbf{B}^c=0$, $\nabla\cdot \mathbf{B}^s=0$, due to absence of magnetic charge and spin-magnetic charge. However, if singular configuration is allowed, the divergences may admit singularities in the bulk and the total contribution in the bulk is quantized due to Dirac quantization condition (or more general Schwinger-Zwanziger quantization condition). For example, let us consider one magnetic monopole (of $A^c_\mu$ gauge group) located at the origin of the three-dimensional space. $\int d^3x \nabla\cdot\mathbf{B}^c=2\pi N_m^c$ where $N_m^c\in \mathbb{Z}$ is an integer-valued ``magnetic charge''. Therefore, the corresponding response total spin $N^s=\int d^3x J^s_0=\frac{\theta_0}{2\pi}N_m^c$ which indicates that  a nonzero Theta term supports a ``{\it polarization spin cloud}'' in the presence of magnetic monopole of $A^c_\mu$ gauge group. A monopole of $A^c_\mu$ gauge group can also trivially attach integer numbers ($n^s_{+}, n^s_{-} \in\mathbb{Z}$)of bosons with $S^z=1,-1$ in the bulk respectively. Therefore, the whole formula of the so-called {\it mutual-Witten effect} can be expressed as:
\begin{align} 
N^s=n^s_{+}-n^s_{-}+\frac{\theta_0}{2\pi}N^c_m\,.\label{3dmutualwitten}
\end{align}
Substituting (\ref{theta01}) into (\ref{3dmutualwitten}) leads to:
\begin{align} 
N^s=n^s_{+}-n^s_{-}+\frac{1}{2}N^c_m\,.\label{3dspinwitten2}
\end{align}
where, $k=0$ is selected for simplicity.  Likewise, we can place a magnetic monopole of $A^s_\mu$ gauge group. Then a ``{\it polarization charge cloud}'' arises. The corresponding mutual-Witten effect is:
\begin{align} 
N^c=n^c+\frac{\theta_0}{2\pi}N^s_m\,.\label{3dmutualwitten1}
\end{align}
where, $n^c$ the number of bosons trivially attached to the magnetic charge of $A^s_\mu$ gauge group, i.e. spin-magnetic charge. 
Substituting (\ref{theta01}) into (\ref{3dmutualwitten1}) leads to:
\begin{align} 
N^c=n^c+\frac{1}{2}N^s_m\,.\label{3dspinwitten3}
\end{align}
\begin{figure}[t]
\centering
\includegraphics[width=8.6cm]{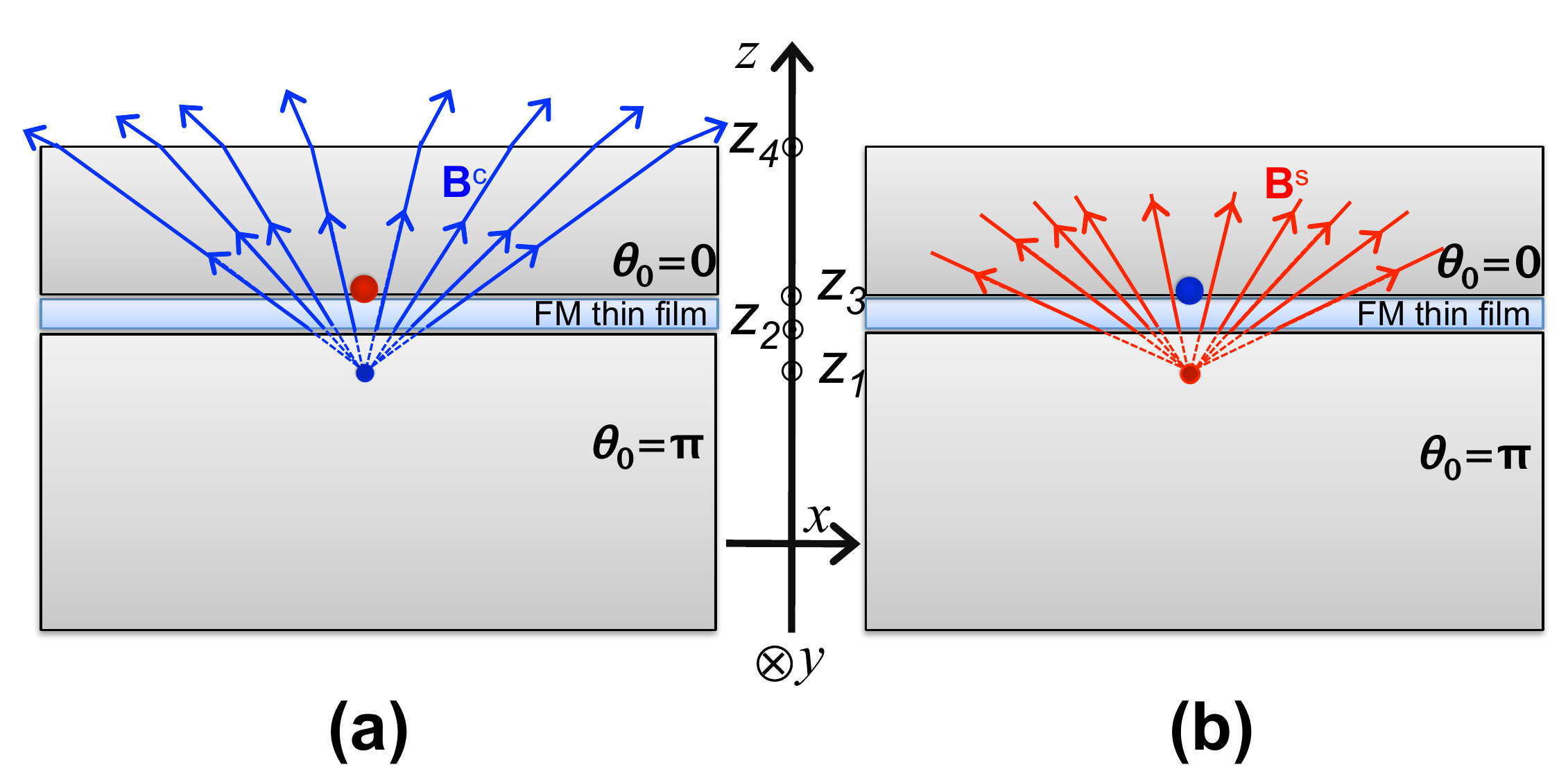}
\caption{(Color online).  Illustration of the experimental setup to realize the mutual-Witten effect. $x$-, $y$-, and $z$-axises form the three spatial directions. In both (a) and (b), a ferromagnetic thin film (FM) is located between a trivial U$_c$(1)$\times$[U$_s$(1)$\rtimes$Z$_2$] state with $\theta_0=0$ and the nontrivial state with $\theta_0=\pi$.  The width of the film is sufficiently small.  In (a), the red and blue balls stand for a spin impurity and an image magnetic monopole of electromagnetic $A^c_\mu$ gauge field, respectively.  The solid blue lines in the region $(z_2, z_4)\bigcup(z_4,\infty)$ represent the magnetic field $\mathbf{B}^c$ induced by the spin impurity. In (b), the blue and red balls stand for an electric charge impurity and an image magnetic monopole of $A^s_\mu$ gauge field (i.e., the ``spin gauge field''), respectively. The solid red lines in the region $(z_2,z_4)$ represent the magnetic field $\mathbf{B}^s$ induced by the electric charge impurity (see texts).}
\label{figure_mutual}
\end{figure}

In Fig. \ref{figure_mutual}, we illustrate the mutual-Witten effect, motivated by Ref. \onlinecite{inducing}. The ferromagnetic (FM) thin film {\it uniformly} breaks Z$_2$ spin rotation symmetry ($\pi$-rotation about spin $S^y$) of the surface of $\theta_0=\pi$ nontrivial bulk. We assume that the width of the FM film is sufficiently small, i.e. $z_3\gtrsim z_2$. In Fig. \ref{figure_mutual}-(a), a spinful but charge-neutral impurity (the red ball) is located at $(x,y,z)=(0,0,z_3)$ near the surface ($z=z_2$). To solve the static electromagnetic problem of both U(1) gauge groups in the region $(z_2,z_4)\bigcup(z_4,\infty)$, one can use the trick of ``image charge''
\cite{Jackson}. One can introduce an image ``spin'' (with spin $N^s$ in unit of $\hbar$) located at $(0,0,z_1)$ with $z_2=(z_1+z_3)/2$ which effectively takes the boundary condition on $z=z_2$ plane into consideration.  At the same position, the {\it mutual Witten effect} admitted by the nontrivial bulk ($z<z_2$) induces an image magnetic monopole (with magnetic charge $N^c_m=2N^s$ in unit of $h/e$, denoted by the blue ball) of $A^c_\mu$ gauge group. This image monopole determines the magnetic field $\mathbf{B}^c$ inside the region $(z_3,z_4)$ (the solid blue lines with arrows). Since these magnetic lines are formed by the electromagnetic field $A^c_\mu$, the magnetic lines can also penetrate $z=z_4$ plane, i.e. the top surface of $\theta_0=0$ bulk, and, flows into the vacuum $(z_4,\infty)$ where both dielectric constant $\epsilon_0$ and permeability $\mu_0$ are nonzero forming nonzero static charge-charge correlation.  The radial magnetic distribution provides a Lorentz force acting on  electrically charged currents. The latter can be excited if temperature is nonzero  inside the trivial  gapped  bulk. It should be noted that the roles of other boundaries (such as $z=z_4$ plane) are ignored since they are irrelevant to the formation of mutual-Witten effect discussed here.

There is one remaining issue to be stressed below. Different from Ref. \onlinecite{inducing} where only electromagnetic field $A^c_\mu$ is considered, we must add a trivial bulk above the FM film. The reason is following. The vacuum medium is really ``empty'' for $A^s_\mu$ gauge field in a sense that the vacuum has no background dynamics admitting communication between two spinful particles (namely, particle-1 with spin $q_1$ and particle-2 with spin $q_2$), which is in contrast to $A^c_\mu$ that has background $\epsilon_0<\infty$ and $\mu_0>0$. In other words, in the vacuum the two spinful particles cannot interact with each other via $A^s_\mu$ gauge field. Classically, it indicates that the ``electromagnetic force'' felt by particle-1 is vanishing: $\mathbf{F}_{12}\equiv q_1 \mathbf{E}^s+q_1\mathbf{v}_1\times \mathbf{B}^s=0$, where, $\mathbf{v}_1$ is the velocity vector of particle-1, and, $\mathbf{E}^s$ and $\mathbf{B}^s$ are spin-electric field and spin-magnetic field formed by particle-2 current.   
On the other hand, the trivial bulk on the top of FM film provides a nontrivial dynamical background where spin-spin correlation is well-formed and thus two spinful particles can talk to each other via $A^s_\mu$ gauge field.

Likewise in Fig. \ref{figure_mutual}-(b), we may place an electric charge impurity (the blue ball) which finally induces a magnetic monopole (the red ball) of $A^s_\mu$ gauge group. The monopole determines the magnetic field $\mathbf{B}^s$ distribution in the trivial state. As explained above, $\mathbf{B}^s$ magnetic lines (the solid red line with arrows) are confined inside the trivial bulk and cannot penetrate into the vacuum, in contrast to $\mathbf{B}^c$.  The radial magnetic distribution provides a Lorentz force acting on  spin currents. The latter can be excited  if temperature is nonzero  inside the trivial  gapped  bulk.

\subsubsection{Anomalous $K_G$-matrix on $\partial\Sigma^3$}

Similar to Sec. \ref{AcKmat},\ref{AsKmat},
let us use the top-down approach explicitly working out the external field theory 
on  $\partial\Sigma^3$ of U$_c$(1)$\times$[U$_s$(1)$\rtimes$Z$_2$] global symmetry with $Z^T_2$ symmetry broken, 
and the external field theory 
on $\Sigma^2$ with U$_c$(1)$\times$U$_s$(1) global symmetry.
Here we save detailed derivations to Appendix \ref{appA} and list down key results directly. 
We firstly study on $\Sigma^2$ with U$_c$(1)$\times$U$_s$(1) global symmetry, what we start with is the intrinsic SPT's $
K_{S} =\bigl( {\begin{smallmatrix} 
0 &1 \\
1 & 0
\end{smallmatrix}}  \bigl) \oplus
\bigl( {\begin{smallmatrix} 
0 &1 \\
1 & 0
\end{smallmatrix}}  \bigl) 
$ and gauging the U(1)$\times$U(1) global symmetry current coupling to $A^{s}$, we obtain $K_{G} ={\begin{pmatrix} 
2 p_1 &p_{12}  \\
p_{12} & 2 p_2 
\end{pmatrix}}$ in the gauge charge sectors of U(1)$\times$U(1). 
In the case of a charge-1 and spin-1 bosonic system, we have spin sum $\sum q_i^2=1$,
\begin{align}
&\cL_{SPT+Gauge}(A^c,A^s) =\nonumber\\
\frac{1}{4\pi}& \epsilon^{\mu\nu\lambda} (A^c_\mu, A^s_\mu)  
{\begin{pmatrix} 
2 p_1 &p_{12}  \\
p_{12} & 2 p_2 
\end{pmatrix}}  \partial_\nu \left( \begin{matrix}A^c_\lambda\\A^s_\lambda\end{matrix} \right) 
\label{scCSKmat}
\end{align}
with $p_1,p_2,p_{12}\in \mathbb{Z}$ labeling the class of the cohomology group $\cH^3(\U(1) \times \U(1) ,\U(1))=\mathbb{Z}^3$.
We comment that this result is more general than Eq. (\ref{u1u1z2CS}) because there U$_c$(1)$\times$[U$_s$(1)$\rtimes$Z$_2$] symmetry restricts
 $\sigma^{c}=\sigma^{s}=0$ so $p_1=0$ and $p_2=0$.
However, the Hall conductance as the response of this $\cL_{SPT+Gauge}(A^c)$ is the same
\be
\sigma^{sc}=\sigma^{cs}=p_{12} \frac{1}{2\pi}  \,,\label{2dCShallKmat}  
\ee
 
On the other hand, as the intrinsic topological field theory of $\Sigma^3$ in 3+1D is not yet known to be completed,  for the anomalous $K_G$-matrix in $\partial\Sigma^3$ 
we simply adopt the result from the previous section to modify the effective $K_G$ matrix on $\partial \Sigma^3$ as 
\be
K_{G,\partial\Sigma^3}={\begin{pmatrix} 
2 p_1 &p_{12}+\frac{\theta_0}{2\pi}  \\
p_{12}+\frac{\theta_0}{2\pi}  & 2 p_2 
\end{pmatrix}}
\ee
in Eq. (\ref{scCSKmat}). With Hall conductance,
\be
\tilde{\sigma}^{sc}=\tilde{\sigma}^{cs}=(p_{12}+\theta_0/2\pi) \frac{1}{2\pi}\,\label{3dSChallKmat}
\ee
Therefore, we have Eq. (\ref{2dCShallKmat}) and Eq. (\ref{3dSChallKmat}) written in a consistent manner as Eq. (\ref{2dhall0}) and Eq. (\ref{3dhall0}). 
The top-down approach here shows the consistency to Sec. \ref{SecIID1}.
 
\section{Discrete Z$_N$ charge symmetry and Z$_N$ spin symmetry}\label{sec:discretesymmetry}
\subsection{Dynamical gauge theory: a general discussion\label{sec:discretesymmetry1}}
In Sec. \ref{sec:continuoussymmetry}, we discussed on the response theory of SPT states with charge / spin continuous symmetry. In the examples we considered, we found exotic $\Sigma^3$ bulk response phenomena which are descendents of original Witten effect although the bulk is fully gapped insulators. We also found exotic $\partial\Sigma^3$ surface Hall effects which contain many variants and all of them cannot be realized in a strictly two-dimensional SPT state with the same symmetry as the $\partial\Sigma^3$. The statement is that, given an SPT with symmetry $G$ in $\Sigma^3$, its surface ($\partial\Sigma^3$) with symmetry $G'$ (as a subgroup of $G$) cannot be  realized in an SPT with the same symmetry $G'$ but defined in $\Sigma^2$.

In this section, we will consider discrete  charge / spin symmetry. Recall that, in a BCS superconductor, the charge symmetry reduces to Z$_2$ from U(1) due to Cooper pairs. As a result, the magnetic flux inside a type-II BCS superconductor is quantized to $\pi$. The charge response current in terms of Ohm's equation is screened in a sense that in the linear response regime the external electromagnetic field cannot be adiabatically turned on from zero due to photon mass. Likewise, in an SPT state with discrete charge / spin  symmetry group which can be acheived by charge-$N$ condensate / spin-$N$ condensate, the response phenomenon in the linear regime is always dominated by Meissner effect.

Instead of the response approach utilized in Sec. \ref{sec:continuoussymmetry}, in the following, we will study these states by gauging the charge / spin symmety which results in the {\it dynamical gauge theory} description of SPT states.  We stress that, all field variables in the dynamical gauge theory description are  now dynamical and  appear in the path-integral measure. Most importantly, this dynamical gauge theory is {\it not} the low-energy theory of the SPT state but a new window / tool to diagnose the SPT states. By studying the dynamical gauge theory, we will find the anomalous surface($\partial\Sigma^3$) - in a sense that the gauged theory on the surface with symmetry $G'$ is different from the gauged theory of a two-dimensional SPT with the same $G'$. And, the dynamical gauge theory bridges SPT to a topological ordered state, namely ``symmetry-enriched topological phase'' (SET) in which the fingerprint of  SPT is hidden.

\subsubsection{Gauging SPT to SET \label{sec:gaugingSPT}}

The more precise statement of gauging the subgroup $G'$ of the full global symmetry group $G$, is that we convert partially the global symmetry group to the gauge symmetry group, with a leftover global symmetry
($\text{Z}_2$ or $\text{Z}^T_2$). In other words, what we really do is converting SPT states to the symmetry enriched topological(SET) states, \emph{i.e.} topologically ordered states enriched with a global symmetry. Let us illuminate this relation as follows.

\noindent
\fbox{Basics of SET}:

We firstly set up the SET 
picture in Ref. \onlinecite{Hung:2012nf,Mesaros:2012yd,2013PhRvB..87j4406E,Lu:2013jqa}.
Following pioneer works\cite{Hung:2012nf,Mesaros:2012yd,2013PhRvB..87j4406E}, let us define
the global symmetry group as $G_s$, the gauge symmetry group as $G_g$. 
This SET picture in Ref. \onlinecite{Hung:2012nf} considers the exact sequence in SET, 
\be
1\to G_g \to PSG \to G_s \to 1 \label{SETexact}
\ee
which says that projective symmetry group, $PSG$, is an extension of the global symmetry group $G_s$ by the gauge symmetry group $G_g$. 
From the exact sequence, $G_g$ is a normal subgroup.
The global symmetry group $G_s$ is regarded as the quotient group $G_s=PSG/G_g$, 
from the full symmetry group PSG mod out a gauge symmetry group $G_g$.  \\

\noindent
\fbox{Promote SPT to SET}: 
We now clarify that in all of our examples, the full symmetry group has the form 
\be
G=G' \rtimes G'' \;\; \text{or}\;\; G=G' \times G'' , \label{Gsemi}
\ee
with $G'$ as the symmetry group being gauged,
and $G''=\text{Z}_2$ (i.e. the $\pi$-rotation about $S^y$) or $G''=\text{Z}^T_2$ is the leftover global symmetry group. This specific form of $G$ implies that $G'$ is always a normal subgroup of $G$.
Thus, the gauging process for all of our five examples in this section (these are
Z$_N$$\rtimes$Z$^T_2$, 
Z$_N$$\times$[U$_s$(1)$\rtimes$Z$_2$], 
U$_c$(1)$\times$[Z$_N$$\rtimes$Z$_2$], 
Z$_{N_1}$$\times$[Z$_{N_2}$$\rtimes$Z$_2$]; see, also, Table \ref{tab:discrete})
guarantee the forms as Eq. (\ref{Gsemi}) corresponding to the Eq. (\ref{SETexact}) in SET picture\cite{Hung:2012nf}.

Let us now tie everything together. In Ref. \onlinecite{Hung:2012nf,2013PhRvB..87j4406E} picture, the full symmetry group $G$ is a projective symmetry group(PSG). Before gauging $G'$, what we have is 
SPT state with $PSG=G_s=G$ and $G_g=\text{Z}_1=1$. After partially gauging the subgroup $G_g=G'$, we can view this as choosing a normal subgroup in the PSG. The leftover global symmetry group
is indeed the quotient group as $G_s=PSG/G_g$.

To be more precise, based on the relation, 
\be
{\text{Gauging SPT: }}  \frac{G}{G'}=G'' \Leftrightarrow {\text{ SET: }} \frac{PSG}{G_g}=G_s,
\ee 
the gauging process for all of our five examples in this section,  can be regarded as converting SPT to SET,\be
\text{SPT} \left\{ \begin{array}{l} PSG=G\\G_g=1\\G_s=\frac{PSG}{G_g}=G \end{array}  \right.  
\xRightarrow{\text{gauging}}
\text{SET} \left\{ \begin{array}{l} PSG=G\\G_g=G'\\G_s=\frac{PSG}{G_g}=\frac{G}{G'} \end{array}   \right.
\ee

In the following and the remaining parts of Sec. \ref{sec:discretesymmetry}, we will base on this principle: 
gauging SPT to SET, try to distinguish the features of SPT from the dynamical gauge theory viewpoint of SET.

\begin{table*}
 \begin{tabular}[t]{|c|c|c|c|c||c|}
 \hline
\begin{minipage}[t]{1in}Full symmetry group PSG \end{minipage} & \begin{minipage}[t]{0.6in}Gauge symmetry group $G_g$ \end{minipage}& \begin{minipage}[t]{0.6in}Global symmetry group $G_s$\end{minipage}&\begin{minipage}[t]{1.4in}3D bulk ($\Sigma^3$) dynamical gauge theory\end{minipage}&\begin{minipage}[t]{1.1in}Surface ($\partial\Sigma^3$) boson theory with anomaly\end{minipage}&
 \begin{minipage}[t]{1.4in} 2D plane ($\Sigma^2$) dynamical gauge theory with $K_G$-matrix\end{minipage}\\
 \hline\hline
 Ê\begin{minipage}[t]{1in}Z$_N$$\rtimes$Z$^T_2$\end{minipage}& \begin{minipage}[t]{0.6in} Z$_N$ \end{minipage}& Ê \begin{minipage}[t]{0.6in} Z$^T_2$ \end{minipage}&\begin{minipage}[t]{1.4in} Ê$\frac{N}{4\pi} \epsilon^{\mu\nu\lambda\rho} B^c_{\mu\nu}\partial_\lambda A_\rho^c+\frac{\theta_c}{8\pi^2} \epsilon^{\mu\nu\lambda\rho}\partial_\mu A^c_\nu \partial_\lambda A^c_\rho$ Ê\end{minipage}&\begin{minipage}[t]{1.1in}
 Z$^T_2$-broken $\partial\Sigma^3$: %
$\frac{N}{2\pi}\partial_0 \phi^c  \epsilon^{ij}\partial_i \lambda^c_j$
\end{minipage}&\begin{minipage}[t]{1.4in}
 Z$^T_2$-broken $\Sigma^2$: %
$ \bigl( {\begin{smallmatrix} 
2p &N \\
N & 0
\end{smallmatrix}}  \bigl)$  
 \end{minipage}\\
\hline
\begin{minipage}[t]{1in}Z$_N$$\times$Z$^T_2$\end{minipage}& Ê\begin{minipage}[t]{0.6in}Z$_N$ \end{minipage}& Ê\begin{minipage}[t]{0.8in} Z$^T_2$\end{minipage}&\begin{minipage}[t]{1.4in} $\frac{N}{4\pi} \epsilon^{\mu\nu\lambda\rho} B^s_{\mu\nu}\partial_\lambda A_\rho^s+\frac{\theta_s}{8\pi^2} \epsilon^{\mu\nu\lambda\rho}\partial_\mu A^s_\nu \partial_\lambda A^s_\rho$ \end{minipage}&\begin{minipage}[t]{1.1in}
 Z$^T_2$-broken $\partial\Sigma^3$:
$\frac{N}{2\pi}\partial_0 \phi^s  \epsilon^{ij}\partial_i \lambda^s_j$
  \end{minipage}&\begin{minipage}[t]{1.4in}
 Z$^T_2$-broken $\Sigma^2$:
$ \bigl( {\begin{smallmatrix} 
2p &N \\
N & 0
\end{smallmatrix}}  \bigl)$
 \end{minipage}\\
\hline
\begin{minipage}[t]{1in}Z$_N$$\times$[U$_s$(1)$\rtimes$Z$_2$]\end{minipage}& Ê\begin{minipage}[t]{0.6in} Z$_N\times$U$_s$(1)\end{minipage}& Ê\begin{minipage}[t]{0.8in}Z$_2$\end{minipage}&\begin{minipage}[t]{1.4in} $\frac{N}{4\pi} \epsilon^{\mu\nu\lambda\rho} B^c_{\mu\nu}\partial_\lambda A_\rho^c+\frac{\theta_0}{4\pi^2} \epsilon^{\mu\nu\lambda\rho}\partial_\mu A^s_\nu \partial_\lambda A^c_\rho$  \end{minipage}&\begin{minipage}[t]{1.1in}
 Z$_2$-broken $\partial\Sigma^3$: 
$\frac{N}{2\pi}\partial_0 \phi^c  \epsilon^{ij}\partial_i \lambda^c_j$

 \end{minipage}&\begin{minipage}[t]{1.4in}
 Z$_2$-broken $\Sigma^2$: %
$ \biggl( {\begin{smallmatrix} 
2p_1 &N & p_{12}& 0\\
N & 0 &0 & 0\\
 p_{12} & 0 &2p_2 & 0\\
0 & 0 &0 & 0
\end{smallmatrix}}  \biggl)$  
 \end{minipage}\\
\hline
\begin{minipage}[t]{1in}U$_c$(1)$\times$[Z$_N$$\rtimes$Z$_2$]\end{minipage}& Ê\begin{minipage}[t]{0.6in}U$_c(1)\times$Z$_N$\end{minipage}&Ê\begin{minipage}[t]{0.8in}Z$_2$\end{minipage}&\begin{minipage}[t]{1.4in} $\frac{N}{4\pi} \epsilon^{\mu\nu\lambda\rho} B^s_{\mu\nu}\partial_\lambda A_\rho^s+\frac{\theta_0}{4\pi^2} \epsilon^{\mu\nu\lambda\rho}\partial_\mu A^s_\nu \partial_\lambda A^c_\rho$  \end{minipage}&\begin{minipage}[t]{1.1in}
 Z$_2$-broken $\partial\Sigma^3$: %
$\frac{N}{2\pi}\partial_0 \phi^s  \epsilon^{ij}\partial_i \lambda^s_j$
 \end{minipage}&\begin{minipage}[t]{1.4in}
 Z$_2$-broken $\Sigma^2$:
$ \biggl( {\begin{smallmatrix} 
2p_1 &0 & p_{12}& 0\\
0 & 0 &0 & 0\\
 p_{12} & 0 &2p_2 & N\\
0 & 0 &N & 0
\end{smallmatrix}}  \biggl)$  
 \end{minipage}\\
\hline
\begin{minipage}[t]{1in}Z$_{N_1}$$\times$[Z$_{N_2}$$\rtimes$Z$_2$]\end{minipage}& Ê\begin{minipage}[t]{0.6in}Z$_{N_1}\times$Z$_{N_2}$\end{minipage}& Ê\begin{minipage}[t]{0.8in} Z$_2$  \end{minipage}&\begin{minipage}[t]{1.4in}$\frac{N_1}{4\pi} \epsilon^{\mu\nu\lambda\rho} B^c_{\mu\nu}\partial_\lambda A_\rho^c+\frac{N_2}{4\pi} \epsilon^{\mu\nu\lambda\rho} B^s_{\mu\nu}\partial_\lambda A_\rho^s+\frac{\theta_0}{4\pi^2} \epsilon^{\mu\nu\lambda\rho}\partial_\mu A^s_\nu \partial_\lambda A^c_\rho$ \end{minipage}&\begin{minipage}[t]{1.1in}
 Z$_2$-broken $\partial\Sigma^3$: 
$\frac{N_1}{2\pi}\partial_0 \phi^c  \epsilon^{ij}\partial_i \lambda^c_j+\frac{N_2}{2\pi}\partial_0 \phi^s  \epsilon^{ij}\partial_i \lambda^s_j$
 \end{minipage}&\begin{minipage}[t]{1.4in}
 Z$_2$-broken $\Sigma^2$:
$ \biggl( {\begin{smallmatrix} 
2p_1 &N_1 & p_{12}& 0\\
N_1 & 0 &0 & 0\\
 p_{12} & 0 &2p_2 & N_2\\
0 & 0 &N_2 & 0
\end{smallmatrix}}  \biggl)$  
 \end{minipage}\\
\hline
\hline
 \end{tabular}
 Ê\caption{The dynamical gauge theory description of spin-1 and charge-1 boson SPT systems with discrete spin symmetry and / or discrete charge symmetry. By following  Sec. \ref{sec:discretesymmetry} and specifically Sec. \ref{sec:gaugingSPT}\label{tab:discrete}, we connect the concept of gauging symmetry-protected topological(SPT) states to symmetry-enriched topological(SET) states. 
The first column: projective symmetry group(PSG) in SET corresponds to the full symmetry group in SPT. 
The second column: $G_g$ in SET corresponds to the gauged symmetry group in SPT.
The third column: $G_s$ in SET corresponds to the remaining ungauged symmetry group in SPT. Z$_2$ in this column is $\pi$-rotation about $S^y$. 
The fourth column shows the effective dynamical gauge theory description of SET(more details in Sec. \ref{sec:BF}), which includes topological BF term and $\Theta$-term $F\wedge F$.
The fifth column shows the surface gapless anomalous boson theory on $G_s$-broken $\partial \Sigma^3$ of $\Sigma^3$ bulk SET (more details in Sec. \ref{sec:surfaceBF})
The last column is filled with the dynamical gauge theory of $G_g$-symmetry SPT 
(but with no $G_s$ symmetry) on intrinsic $\Sigma^2$ surface - by gauging the symmetry $G_g$, to compares with the fifth column $\partial \Sigma^3$.   The central messages are: 1) After gauging a normal subgroup of the symmetry of 3D SPT, we obtain an SET state described by a dynamical gauge theory with a remaining global symmetry;  2) The surface (with Z$_2^T$ or Z$_2$ broken) of 3D SET  is described by a gapless boson matter field with quantum anomaly; 3) After fully gauging an SPT on $\Sigma^2$, we obtained a dynamical gauged field theory described by $K_G$-matrix Chern-Simons theory; 4) The resultant  states on $\partial\Sigma^3$ and $\Sigma^2$ are different although both are two-dimensional space manifold.
Be aware that $B^c$ and $B^s$ are external anti-symmetric 2-form $B^c_{\mu\nu}$ and $B^s_{\mu\nu}$, we should {\it not} misunderstand its meaning mixed with magnetic field $\mathbf{B}^c,\mathbf{B}^s$.
}\end{table*}

\subsubsection{Dynamical BF term in $\Sigma^3$ \label{sec:BF}}
 
In this section, we demonstrate that gauging a discrete Z$_N$ symmetry group  will introduce a new topological term, namely, BF term in which all field variables are dynamical.   

To acheive a Z$_N$ gauge theory, we start with a U(1) gauge theory with a gauge field $A_\mu$ ($A_\mu$ will be replaced by  $A^s_\mu$ and $ A^c_\mu$ later) and  couple $A_\mu$ to charge-$N$ bosonic condensate $\chi$.  Adding a Higgs potential $U(\chi)$, the resultant condensate $\langle\chi\rangle$ (or vacuum expectation value) 
will spontaneously break U(1) symmetry down to {Z}$_N$ symmetry. The Lagrangian for {Z}$_N$ gauge theory\cite{Banks:1989ag} is 
\bea
&&|(\partial  -i N   A)\chi|^2+ U(\chi)+\cdots\nonumber\\
&&=\langle\chi\rangle^2 |(\partial \varphi - N  A )|^2+ U(\langle\chi\rangle)+\cdots\,,
\eea
where, $\cdots$ stands for other terms that already exist in the U(1) gauge theory. $\chi=\langle\chi\rangle e^{i\varphi}$.  Below we show the dual description of this {Z}$_N$ gauge theory is BF theory. 
This argument works in arbitrary spacetime dimension $D$, so let us demonstrate more conveniently in differential form. 
Here $F=dA$ is the 2-form field strength of $A$, with $A$ is the 1-form gauge field. 
While $B$ is another $(D-2)$-form with independent gauge degree of freedom different from $A$.
The trick is dualizing $\varphi$ by introducing a Lagrangian multiplier $B$.
Since $d^2\varphi=0$, let us name $\rho=d\varphi$, we can impose $d \rho=0$ by a Lagrangian multiplier $B$, 
\bea
&&\langle\chi\rangle^2(d\varphi-N A) \wedge *(d\varphi-N  A)\nonumber\\
&=&\langle\chi\rangle^2(\rho-N A) \wedge *(\rho-N  A)+\frac{1}{2\pi} B \wedge d \rho
\eea
In the first line of the above equation,  the integral measure of the path-integral is $\mathscr{D}A\mathscr{D}\varphi$, whille in the second line
is changed to $\mathscr{D}A\mathscr{D}\rho\mathscr{D}B$. By writing $B \wedge d \rho=(-1)^{D-2} dB \wedge  \rho$ 
and redefining field $(\rho-Ne A) \to \rho$, we may directly integrate out the configuration $\mathscr{D}\rho$ rendering:
\be
\frac{N}{2\pi} B \wedge  dA +\frac{(-1)^{(D-1)}}{(4\pi \langle\chi\rangle)^2}  dB \wedge * dB
\ee
where, the first term is the topological BF term while the second term is the Maxwell term for $D-2$-form U(1) gauge field $B$. 
The path integral of this action has the measure $\mathscr{D}A\mathscr{D}B$ since we integrated out $\mathscr{D}\rho$.
We assume that the superfluid density $\langle\chi\rangle$ is sufficiently large forming an ultra-violet energy scale and thus the  Maxwell term which is quadratic becomes irrelevant in the low-energy field theory.

In closing the derivation of BF term as the dual description of the Z$_N$ gauge theory, we need to confirm that the prefactor of the BF term ``$\frac{N}{2\pi}B\wedge d A$'', i.e. $N/2\pi$ is the correct normalization by examing whether the {\it statistical angles} of this BF theory exactly recover the result in the {Z}$_N$ gauge theory or not.

In the famous Kitaev's {Z}$_2$ toric code\cite{Kitaev:1997wr} (as {Z}$_2$ topological order or {Z}$_2$ gauge theory\cite{Wenbook}), there are two kinds of excitations $e$ and $m$ anyons. 
When doing a full winding(or twice exchange) between $e$ and $m$, the $e$-and-$m$ wave function gain a $\pi$ phase - so called  ``statistical angle'' $\pi$.
In general, it is known that a $2\pi/N$ statistical angle can be obtained from doing a full winding (or twice exchange) between certain two excitations of ${Z}_N$ gauge theory. On the other hand, in BF theory, these two excitations are a point-particle with spacetime trajectory described by a one-dimensional worldline ${J}$ which minimally couples to $A$ via $A\wedge *{J}$ and 
a higher dimensional object (such as string or membrane) with spacetime trajectory described by a worldsheet or worldvolume $\Sigma$ which minimally couples to $B$ via $B\wedge *\Sigma$. To test the statistical angle,
we determine the statistical interaction between the two matter field spacetime trajectories ${J}$ and $\Sigma$ by studying the following Lagrangian:
\begin{align}
&\frac{N}{2\pi} B \wedge  dA + A \wedge * J +B \wedge * \Sigma\nonumber\\
=&\frac{N}{2\pi (D-2)!}  \epsilon^{\dots} B_{\dots} \partial_{.} A_{.} d^D x+ A_{\mu} J^\mu d^D x \nonumber\\
&+\frac{1}{(D-2)!} B_{\dots} \Sigma^{\dots}d^D x
\end{align}
Here we skip the apparent indices as $\dots$ abbreviation.
By noticing that the path integral of the above Lagrangian has integral measure $\mathscr{D}A\mathscr{D}B\mathscr{D}{J}\mathscr{D}\Sigma$, we integrate out $\mathscr{D}A\mathscr{D}B$ to deduce the statistical interaction, resulting in a Hopf term\cite{Zee:2003mt}:
\be
\frac{2\pi}{N} J_{.} \frac{\epsilon^{\dots}  \partial_{.}}{\partial^2} \Sigma_{\dots} 
\ee
This is the phase appeared in the exponent of the partition function $\exp[i\frac{2\pi}{N} J_{.} \frac{\epsilon^{\dots}  \partial_{.}}{\partial^2} \Sigma_{\dots} ]$, as 
this $\frac{2\pi}{N}$ factor implies a $\frac{2\pi}{N}$ statistical angle when braiding $J$ around $\Sigma$ by $2\pi$.
 which reassures our $\frac{2\pi}{N}$ normalization is correct.

Let us express the differential form explicitly:
\be
B \wedge  dA=\frac{1}{(D-2)!}{\epsilon}^{\mu_1 \mu_2\dots \mu_D} B_{\mu_1 \dots}   \partial_{\mu_{D-1}} A_{\mu_D} d^D x \,.
\ee
In $\Sigma^3$ bulk (i.e. spacetime dimension $D=4$), the BF term is explicitly expressed by:
\be
\mathcal{L}=\frac{N}{4\pi} \epsilon^{\mu \nu\lambda\rho } B_{\mu \nu}   \partial_{\lambda} A_{\rho}\label{bfterm}
\ee
where, $A_\mu\equiv A^c_\mu$ ($A^s_\mu$) and $B_{\mu\nu}\equiv B^c_{\mu\nu}$ ($B^s_{\mu\nu}$)  if Z$_N$ symmetry originates from U$_c$(1) charge symmetry (U$_s$(1) spin symmetry). The path-integral measure is $\mathscr{D}A\mathscr{D}B$. 

If the Z$_N$ gauge theory is defined on $\Sigma^2$ (i.e. spacetime dimension $D=3$), the BF term reduces to a mutual Chern-Simons term:
\be
\mathcal{L}=\frac{N}{2\pi} \epsilon^{\mu \nu\lambda } A_{\mu }   \partial_{\nu}  {\overline{A}_{\lambda}}\,,\label{mts}
\ee
where, a general $(D-2)$-form gauge field $B$ reduces to a simplest one-form gauge field denoted by ${\overline{A}_\mu}$, and the path-integral measure is $\mathscr{D}A\mathscr{D}{\overline{A}}$.  $A_\mu\equiv A^c_\mu$ ($A^s_\mu$) and ${\overline{A}_\mu}\equiv\overline{A}^c_{\mu}$ ($\overline{A}^s_{\mu}$) if Z$_N$ symmetry originates from U$_c$(1) charge symmetry (U$_s$(1) spin symmetry).

We comment that if we view both $B$ and $A$ field in BF theory as dynamical gauge fields, the overall theory is a {\it dynamical topological field theory} - 
in the sense that the ground state degeneracy of BF theory will depend on the topology of the spatial manifold.
For example, 3+1D BF theory has $N^3$ ground state degeneracy, where $N$ is the 
coefficient `level $N$' of BF term. Another example, 2+1D mutual Chern-Simons theory has $N^2$ ground state degeneracy, where $N$ is the 
coefficient `level $N$' of mutual Chern-Simons term\cite{2004AnPhy.313..497H}.
As a side note, recently BF term has been applied to other contexts with different interpretation in condensed matter physics.\cite{2004AnPhy.313..497H, 2011AnPhy.326.1515C,VS1258}

\subsubsection{Surface theory with quantum anomaly\label{sec:surfaceBF}}
 
 In this subsection, we briefly preview 
 the procedure carried out in the next sections Sec. \ref{subsec:Z_NxZ_2},\ref{subsec:U(1)xZ_NxZ_2},\ref{subsec:Z_NxZ_NxZ_2}
 - the comparison of two kinds of dynamical gauge theory on the anomalous surface $\partial\Sigma^3$ and on the intrinsic $\Sigma^2$ bulk, 
 both have 2D spatial dimensions.
 The philosophy is that we will treat the external fields appeared in Sec. \ref{sec:continuoussymmetry} as dynamical gauge fields - to gauge the 3D bulk $\Sigma^3$ and study its gauged surface theory $\partial \Sigma^2$ 
 (with Z$_2$ or  Z$^T_2$ symmetry broken),
and compare it to the gauged intrinsic 2D bulk $\Sigma^2$ (without Z$_2$ or  Z$^T_2$ symmetry).
 We find the gauged theory on the anomalous surface $\partial\Sigma^3$ with symmetry $G'$ are different from the gauged theory of a two-dimensional SPT with the same $G'$ in
  the intrinsic $\Sigma^2$ bulk.\\

We call this $\partial\Sigma^3$ surface anomalous, because the boundary field theory on $\partial\Sigma^3$ is meant to cancel the anomaly contributed from the dynamical gauge theory in $\Sigma^3$.
Here the 
situation is similar to the case that 2+1D bulk topological Chern-Simons theory requires 1+1D Wess-Zumino-Witten model on the boundary to cancel the anomaly\cite{Elitzur:1989nr,Treiman:1986ep,Wen:1995qn}. 
A more familiar case is 2+1D bulk Abelian Chern-Simons theory requires 1+1D edge theory of chiral bosons, to preserve the gauge-invariance on the manifold with boundary\cite{Wen:1995qn,Wenbook}.
Similarly, the 3+1D bulk topological BF theory requires the 2+1D anomalous edge theory of chiral   bosons, to preserve the gauge-invariance on the manifold with boundary. We comment that the interpretation of 2+1D electromagnetism  in Ref. \onlinecite{VS1258} is improper. Instead, we interpret the surface theory of BF theory as an anomalous chiral boson theory and  leave the details to Appendix \ref{sec:appendix_surface}.
More on the understanding on the anomaly of topological phase or topological field theory,
and their relation to bulk-edge correspondence can be found in Refs.\onlinecite{Kao:1996ey,Wen:2013oza,Wen:2013ue} and in particular Sec VI of Ref.\onlinecite{Wang:2013yta}.

\subsection{Derivation of surface chiral boson theory: An example with  Z$_{N}\rtimes$Z$^T_2$ \label{subsec:Z_NxZ_2}}
  
Physically, an SPT state with Z$_{N}\rtimes$Z$^T_2$  in $\Sigma^3$ can be viewed as a time-reversal-symmetric bosonic superconductor with charge-$N$ bosonic condensate. By collecting Eq. (\ref{bfterm}) and the $\theta_c$-term in Eq. (\ref{three}), we obtain the following dynamical gauge theory with path-integral measure $\mathscr{D}{A^c}\mathscr{D}B^c$ 
\begin{align} 
  \mathcal{L}=&\frac{\theta_c}{8\pi^2}\partial_{\mu}A^{c}_\nu\partial_\lambda A^{c}_\rho\epsilon^{\mu\nu\lambda\rho}+\frac{N}{4\pi} \epsilon^{\mu \nu\lambda\rho } B^c_{\mu \nu}   \partial_{\lambda} A^c_{\rho}\,,
\end{align}  
where, $\theta_c=2\pi+4\pi k$ ($k\in\mathbb{Z}$). 
Its Z$^T_2$-broken surface $\partial \Sigma^3$, 
however, is meant to cancel the anomaly contribution from the bulk topological BF theory.
This derivation is mentioned in Ref. \onlinecite{VS1258} in a different context, but let us still walk through the logic to have coherent discussion. A convenient way to derive the chiral (vector and scalar) bosons is to choose a temporal gauge choice $A^c_0=0$, $B^c_{0i}=0$. The gauge choice itself should not affect overall physics and thus should only base on the convenience.
The equations of motion (EOM) of $A^c_0$ and $B^c_{0i}$ impose the folllowing constraints: $\epsilon^{0ijk}\partial_i B^c_{jk}=0$ and $\epsilon^{0ijk} \partial_j A^c_k=0$ 
which imply $B^c_{jk}=\partial_j \lambda_k-\partial_k \lambda_j $ and $A^c_k=\partial_k \phi$ as pure gauge forms.
One interprets $\lambda_k$ as vector bosons and $\phi$ as a scalar boson.
Let us consider a $\partial \Sigma^3$  formed by $x_1$-$x_2$ (i.e. $x$-$y$) plane at $x_3=0$ (i.e. $z$=0) and then collect the term on $\partial \Sigma^3$ to be
$\frac{1}{2}\frac{N}{4\pi} 4 \int ( B_{12} F^c_{03} + B_{23} F^c_{01} +B_{31} F^c_{02}) =\frac{N}{2\pi}  \int dx_3 \partial_3 (- \lambda_2 F^c_{01}+\lambda_1 F^c_{02})+\dots$,
so the surface theory is described by the action:
\be
\frac{N}{2\pi} \int d^3x \; (\partial_1 \lambda_2 - \partial_2 \lambda_1 )\partial_0  \phi \,.
\ee
By choosing a light-cone gauge\cite{Elitzur:1989nr} $A^c_0+v_1 A^c_1+v_2 A^c_2=0$, we can add velocity\cite{Wen:1995qn} (so the Hamiltonian is not zero) to the boson theory, so the action becomes
\be
\frac{1}{2\pi} \int d^3x\; \epsilon^{ij}\partial_i \lambda_j (N \partial_0   \phi-v_1 \partial_1   \phi -v_2 \partial_2   \phi) \label{bBF}
\ee
with $i,j$ running in $1,2$.
One can massage this surface action into a more symmetric form,
\bea
\frac{1}{4\pi} \int d^3x\; \epsilon^{ij}\partial_i \lambda_j (N \partial_0   \phi-v_1 \partial_1   \phi -v_2 \partial_2   \phi) \nonumber\\
+ \epsilon^{ij} \partial_i   \phi (k \partial_0 \lambda_j  -v_1 \partial_1 \lambda_j   \phi -v_2 \partial_2   \lambda_j) \,.
\eea
The pure gauge forms also affect the $\theta_c$-term on $\partial \Sigma^3$, $\frac{\theta_c}{4\pi^2}\epsilon^{\nu\lambda\rho}  A^{c}_\nu\partial_\lambda A^{c}_\rho=0$ because of $A^c_k=\partial_k \phi$.
So $\theta_c$-term becomes strictly zero on the surface. In this sense Eq. (\ref{bBF}) is the only left-over term, which is required to cancel the anomaly from the bulk BF theory in $ \Sigma^3$.



 For intrinsic Z$_N$ symmetry SPT on Z$_2$-broken $\Sigma^2$,  collecting Eq. (\ref{eq:c2}), Eq. (\ref{2dhall}) and Eq. (\ref{mts}) leads to the dynamical gauge theory:
\begin{align}
\mathcal{L}=&\frac{2p}{4\pi}A^c_\mu \partial_\nu A^c_\lambda \epsilon^{\mu\nu\lambda}+\frac{N}{2\pi} \epsilon^{\mu \nu\lambda}  A^c_{\mu} \partial_\nu {\overline{A}^c_\lambda} \nonumber\\
=&\frac{1}{4\pi}(A^c_\mu,{\overline{A}^c_\mu})\left(\begin{matrix}2p &N \\
N & 0\end{matrix}\right)\partial_\nu\left(\begin{matrix}A^c_\lambda \\{\overline{A}^c_\lambda}\end{matrix}\right)\epsilon^{\mu \nu\lambda} \,
\end{align}
with path-integral measure $\mathscr{D}{A^c}\mathscr{D}{\overline{A}^c}$, integer $p=k$.

Z$_{N}\times$Z$^T_2$ symmetry group is similar and the results are shown in Table \ref{tab:discrete}.
 Derivations of other symmetry groups are straightforward. (Details of derivations can be found in Appendix \ref{sec:appendix_anomaly}).

%
%
%
%
%
%
%

\section{Conclusions}\label{section:conclusion}
 In summary, in this work we study the response theory and dynamical gauge theory approach of bosonic symmetry-protected topological states (SPT) at least with charge symmetry (U(1) or Z$_N$) or spin $S^z$ symmetry (U(1) or Z$_N$) in 2D bulk, 3D bulk, and the surface of 3D bulk. The response theory applied in the case of continuous U(1) spin or charge symmetry group is based on the minimal physical input (such as the principle of gauge invariance, absence of topological order) without relying on lattice microscopic models. The 3D examples contain U$_c$(1)$\rtimes$Z$^T_2$, U$_s$(1)$\times$Z$^T_2$, and, U$_c$(1)$\times$[U$_s$(1)$\rtimes$Z$_2$], where, U$_c$(1) is charge conservation symmetry, U$_s$(1) is spin rotation symmetry about $S^z$, Z$^T_2$ is time-reversal symmetry, and Z$_2$ is specified to the $\pi$-rotation symmetry about spin $S^y$. Z$^T_2$-broken and Z$_2$-broken surfaces are focused. The symmetry implementation in 2D examples is the same as the surfaces of 3D examples.  By studying the 3D bulk response, we define many variants of the celebrated Witten effects, i.e. charge-Witten effect, spin-Witten effect, and mutual-Witten effect. The last one is especially discussed in details which exhibits very exotic experimental phenomenon. Through case-by-case  comparing the quantum Hall effects between the surface and 2D bulk with the symmetry implementation, we emphasize that the surface of 3D SPT is anomalous and its existence requires the existence of an extra spatial dimension. The systematical study on the response theory of these SPT states with simple spin and charge symmetry implementation sheds light on the realistic charge and spin response properties of underlying SPT states which will be possibly synthesized in condensed matter materials or cold-atom experiments in the near future. 

  On the other hand, the dynamical gauge theory description is also studied through the concrete examples at least with discrete Z$_N$ spin symmetry or discrete Z$_N$ charge symmetry. The latter can be viewed as bosonic topological superconductors. The 3D examples contain Z$_N\rtimes$Z$^T_2$ (Z$_N$ is charge symmetry), Z$_N\times$Z$^T_2$ (Z$_N$ is spin $S^z$ symmetry), U$_c$(1)$\times$[Z$_{N}\rtimes$Z$_2$] (Z$_2$ is $\pi$-rotation about spin $S^y$), Z$_N\times$[U$_s$(1)$\rtimes Z_2$], and, Z$_{N_1}\times$[Z$_{N_2}\rtimes$ Z$_2$] (Z$_{N_1}$ and Z$_{N_2}$ are charge and spin symmetries, respectively). Z$^T_2$-broken and Z$_2$-broken surfaces are focused. The symmetry implementation in 2D examples is the same as the surfaces of 3D examples. The dynamical gauge theory in 3D bulk is a topological gauge theory with topological BF term + variant of axionic $\Theta$-term. Its surface theory is gapless boson matter field theory with quantum anomaly. The dynamical gauge theory in 2D bulk is described by multi-component dynamical Chern-Simons gauge theory with $K_G$-matrix coefficient.  By studying the dynamical gauge theory, we explicitly show the connection between an SPT in 3D and a symmetry-enriched topological phase (SET) in 3D through the concrete examples.

There are several open questions to be stressed in the future work.  
\begin{enumerate}
\item {\it Symmetry implementation on the surface.} It will be quite interesting to study different symmetry breaking patterns on the surface other than Z$^T_2$-breaking and Z$_2$-breaking. For U$_c$(1)$\rtimes$Z$^T_2$ and U$_s$(1)$\times$Z$^T_2$ SPT states in 3D, Refs. \onlinecite{VS1258,WS13} have discussed many possible symmetry implementations on the surface based on field theory approach. U$_c$(1)$\times$[U$_s$(1)$\rtimes$Z$_2$] in 3D will be an interesting SPT state by studying different symmetry breaking patterns on the surface. For all discrete groups we considered, their surface anomalous theory will be also interesting to be investigated with other symmetry implementation. 
\item {\it Classification.} The Z$_2$ nature of the Theta angles ($\theta_c,\theta_s,\theta_0$) gives one nontrivial state and one trivial state.  There are more classes within the group cohomology level\cite{Chenlong}, 
and possibly some classes beyond group cohomology\cite{VS1258,2013arXiv1303.4301C,2013arXiv1305.5851F}, especially those with Z$_2^T$ symmetry. In our case where we only consider $\Theta$-term and BF term, it will be interesting to search for complete set of topological terms to obtain more nontrivial states.

%
%

\item {\it Lattice realization and exactly solvable model.}

The microscopic model, lattice model and exact solvable model can help to determine more physical properties.
Several works along this direction can be found in Ref.\,\onlinecite{LevinGu,Burnell13,2013arXiv1303.4301C,2013arXiv1305.5851F,Chen:2012hc,Santos:2013uda,Hung:2012kc}.
It is noteworthy that the Z$_N$, U(1) symmetry of the charges and spins can be implemented as the rotor angles in a quantum rotor model.\cite{Chen:2012hc,Santos:2013uda} 
In particular, the SPT state with Z$_2$ symmetry has been constructed where the Z$_2$ symmetry is realized as the Z$_2$ spin degree of freedom of $\sigma_x,\sigma_z$.\cite{LevinGu}
SPT states with Z$_N$ symmetry have been constructed in Ref.\,\onlinecite{Hung:2012kc}.
For a detailed lattice construction of the SPT edge states with a Z$_N$ symmetry can be found in Ref.\,\onlinecite{Santos:2013uda}.
Apparently, the experimental relevant materials for realizing these SPT states will be mostly significant.
Various (charge, spin, mutual) Witten effects we proposed may shed light on the identification of this materials. 
Overall, further connections from our approach (on the response theory and the dynamical gauge theory) to an explicit lattice/experimental realization will be desirable. 

%
%
\item {\it Numerical simulation.} Our dynamical gauge theory formulation has numerical simulation implications. For example, we can apply the procedure in Ref.\,\onlinecite{LevinGu} by adding dynamical gauge field variables on the link, 
while global symmetry acts on the boson/spin on the sites, and then investigate the subsequent gauged SPT. 
This can also be done by cocycles formulation\cite{Hung:2012dx,Hung:2012nf,Wen:2013ue,2013PhRvB..87l5114H} from group cohomology viewpoint. 
A recent tensor network approach in Ref.\,\onlinecite{2013arXiv1303.4190H} using quantum state renormalization\cite{2009arXiv0912.1651V,2007PhRvL..99l0601L,2008PhRvB..78t5116G,2009JPhA...42X4004C,2008AdPhy..57..143V} is applied in identifying AKLT states in one- and two-dimensions.   Based on the construction of SPT lattice models\cite{LevinGu,Burnell13,2013arXiv1303.4301C,2013arXiv1305.5851F} and further gauging the SPT by adding gauge field variable on the links\cite{LevinGu,Swingle12}, it will be applicable to apply similar numerical simulations to identify the gauged SPT(or SET), and further pin down the original SPT.
\end{enumerate} 

 \section*{Acknowledgements} 

We would like to thank Xiao-Gang Wen and Guifre Vidal's organization of {\it Emergence \& Entanglement II} conference during which part of the work was initiated. We are grateful to the hospitality of the Institute for Advanced Study in Tsinghua University during the 2013 Summer Forum on {\it the Interplay of Symmetry \& Topology in Condensed Matter Physics}. J.W. thanks Dalimil Mazac, Luiz Santos and Davide Gaiotto for helpful discussion. 
This research is supported by NSF Grant No.
DMR-1005541, NSFC 11074140, and NSFC 11274192. (J.W.) 
Research at Perimeter Institute is supported by the Government of Canada through Industry Canada and by the Province of Ontario through the Ministry of Economic Development \& Innovation. (P.Y. and J.W.)


\appendix

\section{$K$-matrix Chern-Simons theory for SPT and derivation of Response Theory \label{appA}} 
Here we derive the detailed $K$-matrix construction for SPT order(symmetry-protected topological order) and its response theory. Motivated by pioneer works\cite{LevinGu,Kmatrix,Chenggu,Hung:2012dx,Hung:2012nf,Hungwan,Lu:2013jqa}, however, we still keep our discussion below self-contained and accessible.
In 2+1D, it is believed that a large class of SPT orders, especially Abelian SPT orders, can be classified and categorized by Abelian $K$-matrix Chern-Simons theory\cite{Wen:1992uk,Wenbook}.
The intrinsic field theory description of SPT has the following action,  
\be
S_{SPT,\Sigma^2}=\int dt\;d^2x \frac{1}{4\pi} K_{S,IJ}\epsilon^{\mu\nu\rho} a^I_\mu \partial_\nu a^J_\rho 
\ee
where $a$ is the intrinsic gauge field(or so called statistical gauge field), and $K_{S}$ is the $K$-matrix which classifies and categorizes the SPT orders. 

The SPT order is symmetry-protected, so by definition its order is protected by global symmetry - say some global symmetry group $G_s$. The
distinct features of SPT from trivial insulator is its boundary edge states. The effective degree of freedom of SPT edges is chiral boson field $\Phi$, where $\Phi$ is
introduced to preserve action invariance on the boundary under gauge transformation of the field $a$\cite{Wenbook}. The boundary action is 
\be
S_{SPT,\partial \Sigma^2}= \frac{1}{4\pi} \int_{} dt \; dx \; K_{S,IJ} \partial_t \Phi_{I} \partial_x \Phi_{J} -V_{IJ}\partial_x \Phi_{I}   \partial_x \Phi_{J} 
\ee
When $G_s$ symmetry is preserved, the SPT edge states are gapless(otherwise it has degenerated ground states when adding symmetry-allowed gapping term). 
The SPT has ground state degeneracy(GSD) on the torus as $\GSD=|\det K|=1$,\cite{Wenbook,Kmatrix,Wang:2012am} this leads to the constrained canonical form of $K_{S}$.
In this paper we focus on the bosonic Abelian SPT. Due to its bosonic statistics, the quadratic form has all even integer coefficient,
the canonical form\cite{Wang:2012am,canonical} is known to be the $K$-matrix $K^{b\pm}_{N\times N}$, composed by blocks of
$ \bigl( {\begin{smallmatrix} 
0 &1 \\ 
1 & 0  
\end{smallmatrix}} \bigl)$ and a set of all positive(or negative) coefficients $\E_8$ lattices
$K_{\E_8}$. We can explicitly write down $K^{b\pm}_{N\times N}$
\be
K^{b+}_{N\times N} =K^{b0}_{}   \oplus  K_{\E_8} \oplus K_{\E_8} \oplus  \dots 
\ee
and 
\be
K^{b-}_{N\times N} =K^{b0}_{}  \oplus  (-K_{\E_8}) \oplus (-K_{\E_8}) \oplus  \dots 
\ee
where 
\be
K^{b0}_{N\times N} =\bigl( {\begin{smallmatrix} 
0 &1 \\
1 & 0
\end{smallmatrix}}  \bigl) \oplus \bigl( {\begin{smallmatrix} 
0 &1 \\
1 & 0 
\end{smallmatrix}}  \bigl) \oplus \dots 
\ee
and 
\be
  K_{\E_8}=   \begin{pmatrix}
      2 & -1 & 0 & 0 & 0 & 0 & 0 & 0\\
      -1 & 2 & -1 & 0 & 0 & 0 & 0 & 0 \\
      0 & -1 & 2 & -1 & 0 & 0 & 0 & -1 \\
      0 & 0 & -1 & 2 & -1 & 0 & 0 & 0 \\
      0 & 0 & 0 & -1 & 2 & -1 & 0 & 0 \\
      0 & 0 & 0 & 0 & -1 & 2 & -1 & 0 \\
      0 & 0 & 0 & 0 & 0 & -1 & 2 & 0 \\
      0 & 0 & -1 & 0 & 0 & 0 & 0 & 2 \\
    \end{pmatrix} 
\ee
In our paper, however, we will not need $K_{\E_8}$ state for our SPT examples.
While it has been discussed in Ref. \onlinecite{Kmatrix} that many classes of SPT can be realized by rank-2 
$K$-matrix, here we will show some SPT examples in our study need to have $K$-matrix of large ranks, such as rank-4. 

The implementation of this global symmetry can be explicitly shown by the symmetry transformation on the chiral bosons of the edge states,
\be
g: \{  W^{g}, \delta \Phi^{g}, \eta_g \}
\ee

The group element $g$ of symmetry group $G_s$ acts on chiral boson fields as 
\bea
\Phi \to \eta_g  {(W^g)}^{-1}\Phi + \delta\Phi^g\\
K \to \eta_g {(W^g)}^T K W^g
\eea
where $\eta_g=\pm1$,  with $+$ for the unitary symmetry and $-$ for the anti-unitary symmetry transformation. 

We will use the structure of the $G_s$ to constraint the allowed $g$ as $\{  W^{g}, \delta \Phi^{g}, \eta_g \}$. 
The constraint is: under any $\prod_i g_i=\mathbf{e}$, we have chiral boson field as a quantum phase unchanged up to module $2\pi$,
\be
\Phi\xrightarrow{\prod_i g_i=\mathbf{e}}\Phi\; \text{mod}\; 2\pi
\ee

We should also allow the gauge equivalence to identify the same phases disguised by seemly different transformation. Those gauge transformation
is represented by some $\{X,\Delta \Phi\}$ where $X$ obeys $X^TKX=K$ and $X \in GL(N,\mathbb{Z})$ as a general linear group of degree $N$ over integer $\mathbb{Z}$, and $\Phi \to \Phi+ \Delta \Phi$, such that we identify:
\bea
&&W_g \to X^{-1} W_g X,\\
&&\delta \Phi^{g} \to X^{-1} (\Delta \Phi+ \delta \Phi^{g}  -\eta_g  W_g^{-1} \Delta \Phi )
\eea

To gauge the theory, we need to couple the global symmetry current to the (dynamical or external) gauge field $A$. In the specific examples below we study
(such as U(1), Z$_N$, U(1) $\times$ U(1), Z$_{N_1}$ $\times$ U(1), Z$_{N_1} \times Z_{N_2}$), all the global symmetry can be restricted to $g$ as 
$g=\{  W^{g}=\mathbb{I}, \delta \Phi^{g}, \eta_g=+1 \}$, so the global symmetry current is fully determined by $\delta \Phi^{g}$.
Therefor the global symmetry current on 1+1D (here $\partial \Sigma_2$) is $\epsilon^{\mu\nu} \partial_\nu \Phi/2\pi$,
coupled to the external gauge field $A$ as
\be
\q^I_J \frac{1}{2\pi} \epsilon^{\mu\nu\rho} A^I_\mu \partial_\nu \Phi^J 
\ee
the global symmetry current in 2+1D (here $\Sigma_2$) is $ \epsilon^{\mu\nu\rho}\partial_\nu a^J_\rho/2\pi$,
coupled to the external gauge field $A$ as
\be
\q^I_J \frac{1}{2\pi} \epsilon^{\mu\nu\rho} A^I_\mu \partial_\nu a^J_\rho 
\ee

The $t^I_J$ vector is proportional to $\delta \Phi^{g}_J$, with $I$th of $t^I_J$ specify the $I$th independent generator of the group $G_s$. 
The Lagrangian of the SPT order with intrinsic $a$ coupled to the external gauge field $A$ will be:
\be
\cL_{SPT+Gauge}=\frac{1}{4\pi} K_{S,IJ}\epsilon^{\mu\nu\rho} a^I_\mu \partial_\nu a^J_\rho +\q^I_J \frac{1}{2\pi} \epsilon^{\mu\nu\rho} A^I_\mu \partial_\nu a^J_\rho 
\ee

To integrate out $a$, we adopt EOM as a constraint:
\bea
&&\frac{1}{2\pi} K_{S,IJ}\epsilon^{\mu\nu\rho}  \partial_\nu a^J_\rho+\q^J_I \frac{1}{2\pi} \epsilon^{\mu\nu\rho}  \partial_\nu A^J_\rho =0 \nonumber\\
&\Rightarrow&a^{I'}_\rho=- K^{-1}_{S,I'I} \q^J_I  A^J_\rho
\eea
We get the gauged version description, left with only external gauge field $A$,
\bea
\cL_{SPT+Gauge}(A) &=&\frac{1}{4\pi} \epsilon^{\mu\nu\rho} A^{I'}_\mu (-\q^{I'}_I K^{-1}_{S,IJ}  \q^{J'}_J) \partial_\nu A^{J'}_\rho \nonumber\\
&\equiv&\frac{1}{4\pi} \epsilon^{\mu\nu\rho} A^{I'}_\mu  K_{G,I'J'}   \partial_\nu A^{J'}_\rho  \\
\text{where}\; K_{G,I'J'} &\equiv& -\q^{I'}_I K^{-1}_{S,IJ}  \q^{J'}_J  \nonumber
\eea
We will work through examples shown in our main text, relevant to the response study of $\Sigma_2$, $\partial \Sigma_3$.
To reiterate the examples below only requires $W^g=\mathbb{I}$ and $\eta_g=+1$, so below we only list down $\delta \Phi$ to specify the symmetry transformation.
Aa a side remark, our group elements $g$ representation also form a faithful representation\cite{Kmatrix,2013PhRvB..87l5114H}.

\subsection{U(1)}
A rank-2 $K$-matrix suffices to exhaust all classes of group cohomology $\cH^3(U(1),U(1))=\mathbb{Z}$ with U(1) symmetry,
\be
K_{SPT} =\bigl( {\begin{smallmatrix} 
0 &1 \\
1 & 0
\end{smallmatrix}}  \bigl) 
\ee
where the symmetry transformation of U(1) with an angle $\theta$ specifies the group element $g$,
\be
g_{\theta}: \delta \Phi^{U(1)_{\theta}} =\theta \q=\theta {\begin{pmatrix} 
1  \\
-p  
\end{pmatrix}}
\ee
Since a U(1) group only requires one generator, there is only one kind of charge vector $\q=(1,p)$. 
Here $p$ labels the $\mathbb{Z}$ class of the cohomology group $\cH^3(U(1),U(1))=\mathbb{Z}$.
While
$K_{G}= -\q^{}_I K^{-1}_{S,IJ}  \q^{}_J=2p$. So the topological term in the gauged theory is
\be
\cL_{SPT+Gauge}(A) =\frac{2p}{4\pi} \epsilon^{\mu\nu\rho} A^{I}_\mu   \partial_\nu A^{I}_\rho
\ee

\subsection{$\text{Z}_N$}
Similarly as U(1) symmetry case, the Z$_N$ symmetry only requires a rank-2 $K$-matrix, which exhaust all $\cH^3(\mathbb{Z}_N,U(1))=\mathbb{Z}_N$  
\be
K_{SPT} =\bigl( {\begin{smallmatrix} 
0 &1 \\
1 & 0
\end{smallmatrix}}  \bigl) 
\ee

\be
g_{n}: 
\delta \Phi^{}= \frac{2\pi}{N} n {\begin{pmatrix} 
1  \\
-p  
\end{pmatrix}}
\ee
Here $p$ labels the $\mathbb{Z}_N$ class of the cohomology group $\cH^3(\mathbb{Z}_N,U(1))=\mathbb{Z}_N$.
Both $p$ and $n$ have module $N$ structure as elements in $\mathbb{Z}_N$.

However, the main difference from U(1) gauged theory is that for $\mathbb{Z}_N$ case, the gauge charge and gauge flux are quantized by module $N$,
which can be captured by a mutual Chern-Simons term $\frac{N}{2\pi} \epsilon^{\mu\nu\rho}  A^{I}_\mu \partial_\nu A^{II}_\rho$ (or more generally a BF theory, see Sec. \ref{sec:BF}),
where the statistics angle of a full wave function gains a $2\pi/N$ phase after a full winding between a unit gauge charge and a unit gauge flux. 
\be
\cL_{SPT+Gauge}(A) =\frac{1}{4\pi} \epsilon^{\mu\nu\rho} A^{I'}_\mu  {\begin{pmatrix} 
2 p & N  \\
N & 0  
\end{pmatrix}}_{I'J'}  \partial_\nu A^{J'}_\rho
\ee

\subsection{U(1) $\times$ U(1) \label{U(1)U(1)}}
We require a rank-4 $K$-matrix to obtain all classes of group cohomology $\cH^3(U(1)\times U(1),U(1))=\mathbb{Z}^3$ with U(1) symmetry,
\be
K_{SPT} =\bigl( {\begin{smallmatrix} 
0 &1 \\
1 & 0
\end{smallmatrix}}  \bigl) \oplus \bigl( {\begin{smallmatrix} 
0 &1 \\
1 & 0 
\end{smallmatrix}}  \bigl)  
\ee
\bea
g_{\theta}: \delta\Phi= \delta \Phi^{U(1)_{\theta_1}} +\delta \Phi^{U(1)_{\theta_2}} =\theta_1 \q^1+ \theta_2 \q^2\\
\text{with}\;\;\;
\q^1= {\begin{pmatrix} 
1  \\
-p_1  \\
0\\
-p_{12}
\end{pmatrix}},\;\;
\q^2= {\begin{pmatrix} 
0  \\
-p_{21}  \\
1\\
-p_2
\end{pmatrix}}
\eea
with $\theta_1,\theta_2\in \U(1)$.


The terms with gauge fields coupling to the symmetry current are
\be
\q^1_J \frac{1}{2\pi} \epsilon^{\mu\nu\rho} A^1_\mu \partial_\nu a^J_\rho +
\q^2_J \frac{1}{2\pi} \epsilon^{\mu\nu\rho} A^3_\mu \partial_\nu a^J_\rho
\ee
Here we couple the two generators of symmetry group to different gauge fields, and purposefully choose them to be $A^1$ and $A^3$ to represent the charge sector\cite{Hung:2012nf,Hungwan} of gauge fields,
while the meaning of this choice will be revealed in the next subsection in Section \ref{Zn1Zn2}.
It is easy to see $p_{12}+p_{21}$ identify the same index from the gauged coupling term $ {(p_{12}+p_{21})} \epsilon^{\mu\nu\rho}  A^{1}_\mu \partial_\nu A^{3}_\rho/ {2\pi}$.
So we may identify $p_{12}+p_{21} \to p_{12}$, with $p_1,p_2,p_{12}$ each labels a $\mathbb{Z}$ in $\mathbb{Z}^3$. 
The gauged theory has this $K$-matrix 
\be
K_G={\begin{pmatrix} 
2 p_1 & 0 & p_{12} & 0 \\
0 & 0 & 0 &  0 \\
p_{12}  & 0 & 2 p_2 & 0 \\
0 & 0  & 0 & 0 
\end{pmatrix}}
\ee
or simply in the gauge charge sectors of U(1) $\times$ U(1):
$K_G={\begin{pmatrix} 
2 p_1 & p_{12}  \\
p_{12} & 2 p_2 
\end{pmatrix}}$.

\subsection{$\text{Z}_{N_1} \times \text{Z}_{N_2}$ \label{Zn1Zn2}}
We require a rank-4 $K$-matrix to obtain all classes of group cohomology $\cH^3(Z_{N_1} \times Z_{N_2},U(1))=\mathbb{Z}_{N_1} \times \mathbb{Z}_{N_2}\times \mathbb{Z}_{\gcd(N_1,N_2)}$ with U(1) symmetry,
\be
K_{SPT} =\bigl( {\begin{smallmatrix} 
0 &1 \\
1 & 0
\end{smallmatrix}}  \bigl) \oplus \bigl( {\begin{smallmatrix} 
0 &1 \\
1 & 0 
\end{smallmatrix}}  \bigl)  
\ee

\bea
g: \delta\Phi= \delta \Phi^{1} +\delta \Phi^{2} =\frac{2\pi}{N_1} n_1 \q^1+ \frac{2\pi}{N_2} n_2\q^2\\
\text{with}\;\;\;
\q^1= {\begin{pmatrix} 
1  \\
-p_1  \\
0\\
-p_{12}
\end{pmatrix}},\;\;
\q^2= {\begin{pmatrix} 
0  \\
-p_{21}  \\
1\\
-p_2
\end{pmatrix}}
\eea
with $n_1\in Z_{N_1} , n_2\in Z_{N_2}$.


Again, $p_{12}+p_{21}$ identify the same index from the gauged coupling term $ {(p_{12}+p_{21})} \epsilon^{\mu\nu\rho}  A^{1}_\mu \partial_\nu A^{3}_\rho/ {2\pi}$.
So we may identify $p_{12}+p_{21} \to p_{12}$.
However, 
the two gauged sectors of ${Z}_{N_1}$ symmetry and ${Z}_{N_2}$ symmetry share the same index $p_{12}$.
Therefore, we should emphasize the number of different topological classes specify by $p_{12}$ is identified by $p_{12}\sim p_{12}+c_1 N_1+c_2 N_2=p_{12}+c_{12} \gcd(N_1,N_2)$ for any integer $c_1,c_2$, there is a corresponding integer $c_{12}$ from {\it the Chinese remainder theorem}. 
This means 
\be
p_{12}=p_{12}\; \text{mod}(\gcd(N_1,N_2))
\ee 
Altogether we have $p_1,p_2,p_{12}$ each labels $\mathbb{Z}_{N_1}$, $\mathbb{Z}_{N_2}$, $\mathbb{Z}_{\gcd(N_1,N_2)}$. 
While our argument is based on the symmetry transformation from SPT side, this relation can also be confirmed by a different argument from the statistics angle\cite{Chenggu} of the gauged theory side.
The main difference from $U(1) \times U(1)$ gauged theory is that for Z$_{N_1} \times Z_{N_2}$ case, the gauge charge and gauge flux are quantized by module $N$,
which can be captured by two mutual Chern-Simons terms $\frac{N_1}{2\pi} \epsilon^{\mu\nu\rho}  A^{1}_\mu \partial_\nu A^{2}_\rho+\frac{N_2}{2\pi} \epsilon^{\mu\nu\rho}  A^{3}_\mu \partial_\nu A^{4}_\rho$,
where the statistics angle of a full wave function gains a $2\pi/N_1$(or $2\pi/N_2$) phase after a full winding between a unit gauge charge and a unit gauge flux of Z$_{N_1}$ symmetry(or Z$_{N_2}$ symmetry). 
The gauged theory has this $K$-matrix, 
\be
K_G={\begin{pmatrix} 
2 p_1 & N_1 & p_{12} & 0 \\
N_1 & 0 & 0 &  0 \\
p_{12}  & 0 & 2 p_2 & N_2 \\
0 & 0  & N_2 & 0 
\end{pmatrix}}
\ee


\subsection{$\text{Z}_{N_1} \times \U(1)$ \label{Zn1U1}}

Similar to Sec. \ref{U(1)U(1)},\ref{Zn1Zn2}, we require a rank-4 $K$-matrix to obtain all classes of group cohomology $\cH^3(Z_{N_1} \times \U(1) )=\mathbb{Z}_{N_1} \times \mathbb{Z}\times \mathbb{Z}_{N_1}$ with U(1) symmetry,
\be
K_{SPT} =\bigl( {\begin{smallmatrix} 
0 &1 \\
1 & 0
\end{smallmatrix}}  \bigl) \oplus \bigl( {\begin{smallmatrix} 
0 &1 \\
1 & 0 
\end{smallmatrix}}  \bigl)  
\ee
\bea
g: \delta\Phi= \delta \Phi^{1} +\delta \Phi^{2} =\frac{2\pi}{N_1}n_1 \q^1+ \theta_2 \q^2\\
\text{with}\;\;\;
\q^1= {\begin{pmatrix} 
1  \\
-p_1  \\
0\\
-p_{12}
\end{pmatrix}},\;\;
\q^2= {\begin{pmatrix} 
0  \\
-p_{21}  \\
1\\
-p_2
\end{pmatrix}}
\eea
with $n_1\in Z_{N_1}$ and $\theta_2 \in \U(1)$.

While the $K_G$ matrix from the response theory is derived in the same manner as Sec. \ref{U(1)U(1)},\ref{Zn1Zn2}, so we skip details and directly list down the result
\be
K_G={\begin{pmatrix} 
2 p_1 & N_1 & p_{12} & 0 \\
N_1 & 0 & 0 &  0 \\
p_{12}  & 0 & 2 p_2 & 0 \\
0 & 0  & 0 & 0 
\end{pmatrix}}
\ee
The classification follows the logic in Sec. \ref{U(1)U(1)},\ref{Zn1Zn2}, we have $p_1,p_2,p_{12}$ each labels $\mathbb{Z}_{N1}$, $\mathbb{Z}$, $\mathbb{Z}_{N_1}$.

\section{Surface of 3+1D BF theory is not gauge-invariant}\label{sec:appendix_surface}
 
   
     There are some discussions on the surface theory of BF theory in the literature in different contexts. For example, in Appendix A of Ref. \onlinecite{VS1258}, the surface theory is interpreted as a 2+1D electromagnetism, \emph{i.e.} free photon theory. We comment that the theory is not a theory of photons claimed in the reference. In other words, it is not a usual electromagnetism. The reason is the following. If we use the definitions of ``electric field'' and ``magnetic field'' in the reference, the Lagrangian in Eq. (A6)  can be reexpressed in the form of $\mathcal{L}=\cdots+E^2+B^2$ (in Euclidean metric), where ``$\cdots$'' is an extra term that is not a gauge-invariant term. \emph{Gauge invariance} is a fundamental requirement of electromagnetism. If one insists on the term ``electromagnetism'', it is more appropriate to call it ``anomalous electromagnetism'' which can not exist alone in 2+1D. It can exist as a surface of a 3+1D bulk.

We emphasize that both Lagrangian  in their Eq.(A6) and Lagrangians of the chiral boson theory in our paper are not gauge invariant \emph{alone} on the surface. A gauge invariant theory must be composed by the bulk BF theory and the surface theory \emph{as a whole}.
Thus, a full quantum effect of anomaly of leaking currents to the bulk (here relation between chiral boson and  BF theory is like the relation between chiral boson and Chern-Simons theory in 2+1D quantum hall effects) does happen.

%
%
 
\section{Derivation of Surface chiral boson theory with quantum anomaly}\label{sec:appendix_anomaly}

\subsection{U$_c$(1)$\times$[Z$_{N}\rtimes$Z$_2$] and Z$_{N}\times$[U$_s$(1)$\rtimes$Z$_2$] symmetry groups\label{subsec:U(1)xZ_NxZ_2}}
  Physically, an SPT state with U$_c$(1)$\times$[Z$_{N}\rtimes$Z$_2$] in $\Sigma^3$ can be viewed as a three-dimensional interacting bosonic ground state with charge-1 and spin-1 of discrete spin symmetry Z$_{N}\rtimes$Z$_2$ where Z$_N$ is $2\pi/N$-rotation  about spin-$z$-direction and Z$_2$ is $\pi$-rotation  about  spin-$y$-direction. By collecting Eq. (\ref{bfterm}) and the $\theta_0$-term in Eq. (\ref{three}), we obtain the following dynamical gauge theory with path-integral measure $\mathscr{D}{A^c}\mathscr{D}A^s \mathscr{D}B^s$ 
\begin{align} 
  \mathcal{L}=&\frac{\theta_0}{4\pi^2}\partial_{\mu}A^{c}_\nu\partial_\lambda A^{s}_\rho\epsilon^{\mu\nu\lambda\rho}+\frac{N}{4\pi} \epsilon^{\mu \nu\lambda\rho } B^s_{\mu \nu}   \partial_{\lambda} A^s_{\rho}\,,
\end{align}  
where, $\theta_0=\pi+2\pi k$ ($k\in\mathbb{Z}$).

Similar to the previous Sec. \ref{subsec:Z_NxZ_2}, the 3+1D bulk topological BF theory requires the 2+1D edge theory of chiral (vector and scalar) bosons, to preserve the gauge-invariance on the manifold with boundary\cite{Elitzur:1989nr,Wen:1995qn}.
We again firstly choose a temporal gauge choice $A^s_0=0$, $B^s_{0i}=0$. The gauge choice itself should not affect overall physics, this choice should base only on the convenience.
The EOM of $A^s_0$ and $B^s_{0i}$ impose the following constraints: $\epsilon^{0ijk}\partial_i B^s_{jk}=0$ and $\epsilon^{0ijk} \partial_j A^s_k=0$ which  imply $B^s_{jk}=\partial_j \lambda_k-\partial_k \lambda_j $ and $A^s_k=\partial_k \phi$ as pure gauge forms.
One interprets $\lambda_k$ as vector bosons and $\phi$ as a scalar boson. Let us consider a $\partial \Sigma^3$ formed by $x_1$-$x_2$ plane at $x_3=0$ and then  collect the term on $\partial \Sigma^3$ to be 
\be
\frac{N}{2\pi} \int d^3x \; (\partial_1 \lambda_2 - \partial_2 \lambda_1 )\partial_0  \phi \,.
\ee
By choosing a light-cone gauge $A^c_0+v_1 A^c_1+v_2 A^c_2=0$, the action becomes
\be
\frac{1}{2\pi} \int d^3x\; \epsilon^{ij}\partial_i \lambda_j (N \partial_0   \phi-v_1 \partial_1   \phi -v_2 \partial_2   \phi) \label{bBF2}
\ee
with $i,j$ running in $1,2$.
The pure gauge forms also affect the $\theta_0$-term on $\partial \Sigma^3$, with $\frac{\theta_0}{4\pi^2}\epsilon^{\nu\lambda\rho}  A^{c}_\nu\partial_\lambda A^{s}_\rho=0$ because of $A^s_k=\partial_k \phi$.
So $\theta_0$-term becomes strictly zero on the surface. In this sense Eq. (\ref{bBF2}) is the only left-over term, which is required to cancel the anomaly from the bulk BF theory in $ \Sigma^3$.

%

On Z$_2$-broken $\Sigma^2$,  collecting Eq. (\ref{scCSKmat}) and Eq. (\ref{mts}) leads to the dynamical gauge theory:
\begin{align}
\mathcal{L}=\frac{1}{4\pi}\left(\begin{smallmatrix} A^c_\mu,{\overline{A}^c_\mu},A^s_\mu,{\overline{A}^s_\mu}\end{smallmatrix}\right)\left(\begin{smallmatrix}2p_1 &0& p_{12}&0 \\
0 &0&0 & 0\\p_{12} &0& 2p_2&N \\0 &0&N & 0\end{smallmatrix}\right)\partial_\nu\left(\begin{smallmatrix}A^c_\lambda\\{\overline{A}^c_\lambda}\\A^s_\lambda \\{\overline{A}^s_\lambda}\end{smallmatrix}\right) \epsilon^{\mu\nu\lambda}
\end{align}
with path-integral measure $\mathscr{D}{A^c}\mathscr{D}{\overline{A}^c}$ and $p_1,p_{12},p_2\in\mathbb{Z}$. Z$_{N}\times$[U$_s$(1)$\rtimes$Z$_2$] symmetry group is similar and the results are shown in Table. \ref{tab:discrete}.

   \subsection{Z$_{N_1}\times$[Z$_{N_2}\rtimes$Z$_2$] symmetry group \label{subsec:Z_NxZ_NxZ_2} }

  Physically, an SET state with Z$_{N_1}\times$[Z$_{N_2}\rtimes$Z$_2$]  in $\Sigma^3$ can be viewed as a three-dimensional bosonic superconductor with charge-$N_1$ condensate and spin-1 of discrete spin symmetry Z$_{N}\rtimes$Z$_2$ where Z$_2$ is $\pi$-rotation about $S^y$. By collecting Eq. (\ref{bfterm}) and the $\theta_0$-term in Eq. (\ref{three}), we obtain the following dynamical gauge theory with path-integral measure $\mathscr{D}{A^c}\mathscr{D}A^s\mathscr{D}B^c\mathscr{D}B^s$ 
\begin{align} 
  \mathcal{L}=&\frac{\theta_0}{4\pi^2}\partial_{\mu}A^{c}_\nu\partial_\lambda A^{s}_\rho\epsilon^{\mu\nu\lambda\rho}+\frac{N_1}{4\pi} \epsilon^{\mu \nu\lambda\rho } B^c_{\mu \nu}   \partial_{\lambda} A^c_{\rho}\nonumber\\
  &+\frac{N_2}{4\pi} \epsilon^{\mu \nu\lambda\rho } B^s_{\mu \nu}   \partial_{\lambda} A^s_{\rho}\,,
\end{align}  
where, $\theta_0=\pi+2\pi k$ ($k\in\mathbb{Z}$).

Similar to the previous Sec. \ref{subsec:Z_NxZ_2},\ref{subsec:U(1)xZ_NxZ_2}, the 3+1D bulk topological BF theory requires the 2+1D edge theory of chiral (vector and scalar) bosons, to preserve the gauge-invariance on the manifold with boundary\cite{Elitzur:1989nr,Wen:1995qn,Wenbook}.
We again firstly choose a temporal gauge choice\cite{Elitzur:1989nr,Wen:1995qn,Wenbook,VS1258} $A^c_0=A^s_0=0$, $B^c_{0i}=B^s_{0i}=0$. 
The gauge choice itself should not affect overall physics, this choice should base only on the convenience.
The EOM of $A^c_0,A^s_0,B^c_{0i},B^s_{0i}$ impose the following constraints: $\epsilon^{0ijk}\partial_i B^c_{jk}=\epsilon^{0ijk}\partial_i B^s_{jk}=0$ 
and $\epsilon^{0ijk} \partial_j A^c_k=\epsilon^{0ijk} \partial_j A^s_k=0$. 
These imply $B^c_{jk}=\partial_j \lambda^c_k-\partial_k \lambda^c_j $, $B^s_{jk}=\partial_j \lambda^s_k-\partial_k \lambda^s_j $, $A^c_k=\partial_k \phi^c$ and $A^s_k=\partial_k \phi^s$ as pure gauge forms.
One interprets $\lambda^c_k,\lambda^s_k$ as vector bosons and $\phi^c,\phi^s$ as scalar bosons.
Let us consider $\partial \Sigma^3$ formed by $x_1$-$x_2$ plane at $x_3=0$ and then collect the term on $\partial \Sigma^3$ to be 
\be
 \int d^3x \left(\frac{N_1}{2\pi}(\partial_1 \lambda^c_2 - \partial_2 \lambda^c_1 )\partial_0  \phi^c +\frac{N_2}{2\pi}(\partial_1 \lambda^s_2 - \partial_2 \lambda^s_1 )\partial_0  \phi^s \right)\,.
\ee
By choosing light-cone gauges $A^c_0+v_1 A^c_1+v_2 A^c_2=0$ and $A^s_0+v_1 A^s_1+v_2 A^s_2=0$, the action becomes
\bea
\frac{1}{2\pi} \int d^3x\; \epsilon^{ij}\bigg( \partial_i \lambda^c_j (N_1 \partial_0   \phi^c-v_1 \partial_1   \phi^c -v_2 \partial_2   \phi^c) +\nonumber\\
\partial_i \lambda^s_j (N_2 \partial_0   \phi^s-v_1 \partial_1   \phi^s -v_2 \partial_2   \phi^s)  \bigg)\label{bBF3}
\eea
with $i,j$ running in $1,2$.
The pure gauge forms also affect the $\theta_0$-term on $\partial \Sigma^3$, with $\frac{\theta_0}{4\pi^2}\epsilon^{\nu\lambda\rho}  A^{c}_\nu\partial_\lambda A^{s}_\rho=0$ because of $A^c_k=\partial_k \phi^c$, $A^s_k=\partial_k \phi^s$.
So $\theta_0$-term becomes strictly zero on the surface. In this sense Eq. (\ref{bBF3}) is the only left-over term, which is required to cancel the anomaly from the bulk BF theory in $ \Sigma^3$.


On Z$_2$-broken $\Sigma^2$,  collecting Eq. (\ref{scCSKmat}) and Eq. (\ref{mts}) leads to the dynamical gauge theory:
\begin{align}
\mathcal{L}=\frac{1}{4\pi}\left(\begin{smallmatrix}A^c_\mu,{\overline{A}^c_\mu},A^s_\mu,{\overline{A}^s_\mu}\end{smallmatrix}\right)\left(\begin{smallmatrix}2p_1 &N_1& p_{12}&0 \\
N_1 &0&0 & 0\\p_{12} &0& 2p_2&N_2 \\0 &0&N_2 & 0\end{smallmatrix}\right)\partial_\nu\left(\begin{smallmatrix}A^c_\lambda\\{\overline{A}^c_\lambda}\\A^s_\lambda \\{\overline{A}^s_\lambda}\end{smallmatrix}\right) \epsilon^{\mu\nu\lambda}
\end{align}
with path-integral measure $\mathscr{D}{A^c}\mathscr{D}{\overline{A}^c}\mathscr{D}{A^s}\mathscr{D}{\overline{A}^s}$ and $p_1,p_{12},p_2\in\mathbb{Z}$.


\end{document}